\documentclass[aps,prx,reprint,twocolumn,amsmath,amssymb,floatfix,superscriptaddress,nofootinbib,preprintnumbers]{revtex4-2}
\usepackage{graphicx,comment,color,tikz}

\usepackage{physics}
\usepackage{amsmath}
\usepackage{natbib}
\usepackage{xcolor}
\usepackage[normalem]{ulem}
\usepackage{bbm}
\usepackage{multirow}

\usepackage{pifont}
\newcommand{\cmark}{\ding{51}}%
\newcommand{\xmark}{\ding{55}}%

\usepackage{wasysym} 
\definecolor{dodgerblue}{HTML}{1E90FF}
\usepackage[colorlinks=true, citecolor=dodgerblue,linkcolor=dodgerblue,urlcolor=dodgerblue,pdftitle={RHPE}]{hyperref}

\newcommand{\edit}[1]{#1}
\newcommand{\Var}{\mathrm{Var}}

\newcommand{\poly}{\mathrm{poly}}

\begin{document}

\preprint{MIT-CTP/6008}

\title{Unified Probe of Quantum Chaos and Ergodicity from Hamiltonian Learning}

\author{Nik O. Gjonbalaj}
\email{nikgjonbalaj@g.harvard.edu}
\affiliation{Department of Physics, Harvard University, Cambridge, MA 02138, USA}
\author{Christian Kokail}
\affiliation{Department of Physics, Harvard University, Cambridge, MA 02138, USA}
\affiliation{ITAMP, Harvard-Smithsonian Center for Astrophysics, Cambridge, MA, 02138, USA}
\affiliation{QuEra Computing Inc., 1284 Soldiers Field Road, Boston, MA, 02135, USA}
\author{Susanne F. Yelin}
\affiliation{Department of Physics, Harvard University, Cambridge, MA 02138, USA}
\author{Soonwon Choi}
\affiliation{Center for Theoretical Physics—a Leinweber Institute, Massachusetts Institute of Technology, Cambridge, MA 02139, USA}

\date{\today}

\begin{abstract}
    Developing measures of quantum ergodicity and chaos stands as a foundational task in the study of quantum many-body systems.
    In this work, we propose metrics for these effects based on Hamiltonian learning that unify multiple advantages of existing metrics.
    In particular, we show how ergodicity and chaos improve the robustness of Hamiltonian learning to small errors and furthermore demonstrate that this robustness can be used as a metric for such phenomena.
    We analytically and numerically show that our metrics not only distinguish between integrable and ergodic regimes in various spin chains but also quantify chaos and ergodicity, allowing us to locate regions of parameter space displaying maximal ergodicity and maximal sensitivity to local perturbations.
    Our approach not only provides conceptual ways to study quantum chaos and ergodicity but also presents viable experimental methods for quantum simulators.
\end{abstract}

\maketitle

\section{Introduction}

Understanding chaos and ergodicity in strongly-interacting quantum systems remains an outstanding challenge in quantum many-body physics.
While these phenomena admit precise formulations in classical systems, establishing analogous concepts and experimentally relevant diagnostics in the quantum regime has proven considerably more difficult.
The emergence of programmable quantum hardware is now providing conceptually new routes to address these questions in regimes inaccessible to classical computation, enabling the assembly and control of large-scale interacting quantum systems \cite{Bernien_2017,Bentsen_2019,Bluvstein_2021}.
These platforms offer new opportunities to probe thermalizing dynamics \cite{Kaufman_2016,Schreiber_2015,andersen2025thermalization} and have revealed striking departures from conventional expectations, such as the discovery of quantum many-body scars \cite{Bernien_2017,Turner_2018,Turner_2018_2}, or new universal phenomena such as deep thermalization and Hilbert space ergodicity~\cite{Choi_2023,Cotler_2023,Mark_2024,Shaw_2025}.
Fully leveraging these capabilities demands efficient, experimentally accessible diagnostics that characterize ergodicity and quantum chaos in a unified manner.

\begin{figure}[h!]
    \centering
    \includegraphics[width=0.9
\linewidth]{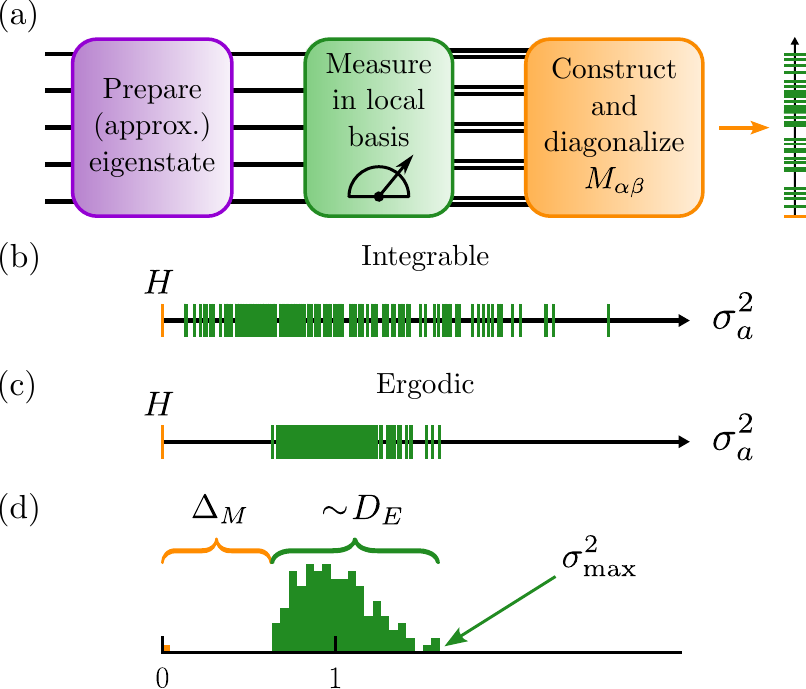}
    \caption{
    \textbf{Ergodicity and Chaos from Hamiltonian Learning.}
    (a) We outline the general procedure for extracting our metrics for ergodicity and chaos. After preparing an (approximate) infinite-temperature eigenstate $\ket{v}$ of a Hamiltonian $H$, local basis measurements are used via e.g. shadow tomography \cite{Huang_2020} to construct the covariance matrix $M_{\alpha \beta}$ [Eq.~\eqref{eq:cov matrix}]. The eigenvalues $\sigma_a^2$ of $M$ constitute the variance spectrum of $\ket{v}$.
    $H$ is learned from the zero-eigenvalue subspace of $M$, while the rest of the variance spectrum acts as an indicator for ergodicity and chaos.
    (b) In integrable systems, the variance spectrum shows a wide spread and a small gap.
    (c) In contrast, ergodic systems are characterized by a narrow variance spectrum with a large gap. 
    (d) We use these metrics (the spread $D_E$ and gap $\Delta_M$) to quantify ergodicity, while the maximum eigenvalue $\sigma^2_{\mathrm{max}}$ probes chaos by quantifying the maximum sensitivity of the eigenstate to a set of local perturbations.
    }
    \label{fig:figure 1}
\end{figure}

\begin{table*}[t!] 
    \centering

    \begin{tabular}{c|c|c|c|c|l} 
         \multicolumn{2}{c|}{Eigenstate-Based} & Level spacing \cite{Oganesyan_2007,Pal_2010,Atas_2013,Atas_2013_2} & $S$ statistics \cite{Rodriguez-Nieva_2024} & AGP norm \cite{Pandey_2020,LeBlond_2021,Kim_2025} & Variance spectrum [this work] \\ 
         \multicolumn{2}{c|}{Ergodicity Metrics}  & $\langle r \rangle$ & $D_{\rm KL}$ & $|| \mathcal{A}_{\lambda}(\mu = N/d)||^2$ & \qquad \quad $\sigma_a^2$ \\
        \hline 
        \hline 
        \multirow{3}{*}{\rotatebox[origin=c]{90}{Properties}} & Quantifies ergodicity? & \xmark & \cmark & \textbf{?} & \qquad \quad \cmark \quad [Fig.~\ref{fig:ergodicity maps}] \\[3pt]
        % \hline
        \cline{2-6}
        & Quantifies eigenstate sensitivity? & \xmark & \xmark & \cmark & \qquad \quad \cmark \quad [Fig.~\ref{fig:chaos map}] \\[3pt]
        & Detects maximal chaos? & \xmark & \xmark & \cmark & \qquad \quad \xmark \quad [Fig.~\ref{fig:chaos map}] \\[3pt]
        \hline
        \multirow{3}{*}{\rotatebox[origin=c]{90}{Meas.}}& Allows approximate eigenstates? & \xmark & \cmark & \cmark & \qquad \quad \cmark \quad [Fig.~\ref{fig:eigenstate preparation}]  \\[5pt]
        & $\ell$-local observable? & \xmark & \xmark & \xmark & \qquad \quad \cmark \quad [Eq.~\eqref{eq:cov matrix}]  \\[5pt]
        \hline 
    \end{tabular}

    \caption{
    \textbf{Measuring Ergodicity from Midspectrum Eigenstates.} 
    We compare various metrics to characterize ergodicity and chaos in quantum many-body systems based on what properties they detect and how they are measured.
    All methods considered use multiple copies of (approximate) energy eigenstates of a \edit{geometrically} $\ell$-local Hamiltonian as the resource for evaluating the metrics. 
    Question marks indicate that it is unknown how to characterize a given property (row) from a metric (column) to the best of our knowledge.
    We note that the AGP norm can be defined for an infinite family parametrized by the energy cutoff $\mu$. In this table, we focus on $\mu = N/d$ \cite{Pandey_2020} for system size $N$ in the Hilbert space of dimension $d = 2^N$ to contrast it with the variance spectrum, but $\mu$ can be tuned to enable estimating the AGP from local observables at the cost of losing its ability to detect maximal chaos.
    Metrics that quantify ergodicity are able to locate ``maximally ergodic'' pockets within the ergodic region of parameter space \cite{Rodriguez-Nieva_2024}.
    While some metrics quantify the sensitivity of eigenstates against local perturbations in $H$, the detection of maximal chaos~\cite{Pandey_2020,LeBlond_2021,Kim_2025}---defined as the sensitivity growing as fast as possible in $N$---requires probing nonlocal responses to the perturbation, as enabled by the AGP norm for $\mu = N/d$.
    Furthermore, some metrics still function when approximate (low energy variance) eigenstates are provided.
    }
    \label{tab:metric comparison}
\end{table*}

Existing diagnostic tools for quantum chaos and ergodicity often rely on spectral properties of the Hamiltonian, using energy eigenvalues and eigenstates to probe aspects of ergodicity.
These approaches build on the predictions of random matrix theory (RMT) and the eigenstate thermalization hypothesis (ETH), which assert that midspectrum eigenstates display universal spectral statistics and locally resemble Gibbs ensembles \cite{Deutsch_1991,Srednicki_1994,D_Alessio_2016,Deutsch_2018}.
Leveraging this framework, a wide range of indicators has been developed, including level-spacing statistics \cite{Oganesyan_2007,Pal_2010,Atas_2013,Atas_2013_2}, spectral form factors \cite{Bertini_2018,Gharibyan_2018,Chan_2018,Friedman_2019,Garratt_2021}, eigenstate entanglement \cite{Vidmar_2017,Lu_2019,Murthy_2019,Lydzba_2020,Kumari_2022,Bianchi_2022,Rodriguez-Nieva_2024,Langlett_2024,Langlett_2025}, and adiabatic deformations \cite{Pandey_2020,LeBlond_2021,Kim_2025,Kim_2024,Karve_2025}. 
However, each of these metrics has its own pros and cons (see Table~\ref{tab:metric comparison}). 
Some require access to operators with extensive support and thus nonlocal measurements or deep circuits to evaluate. 
Others diagnose only ergodicity in the ETH sense, while those that probe other aspects of quantum chaos---such as the susceptibility of eigenstates to perturbations \cite{Kim_2025}---depend on specifying a relevant perturbation of interest. 
Finally, some metrics break down when applied to approximate eigenstates with small but finite energy variance, which is an important consideration for experimental implementation.

In this work, we address the problem of measuring chaos and ergodicity from a learning-theoretic point of view.
Our perspective is motivated by Hamiltonian learning protocols that reconstruct local Hamiltonians from single eigenstates -- techniques originally developed for verification and characterization of quantum devices \cite{Swingle_2014,Garrison_2018,Chertkov_2018,Qi_2019,Lindner-Bairey_2019,Bentsen_2019,evans2019scalablebayesianhamiltonianlearning,Hsieh-Li_2020,Dutkiewicz_2024,Hu_2025}. 
Here, however, we use the learning problem itself as a probe of quantum chaos and ergodicity. Specifically, we ask: \textit{Does the robustness of inferring a parent Hamiltonian from a single midspectrum eigenstate---i.e., its stability under small perturbations or noise---depend on whether the underlying system is integrable or chaotic?} 
We answer this question in the affirmative via detailed analytical and numerical investigations, showing that chaos and ergodicity make Hamiltonian learning \textit{easier} by enhancing its robustness to errors. 
Most importantly, we demonstrate that this effect is strong enough that ergodicity and chaos can be \textit{measured} directly through the robustness of Hamiltonian learning.

Our approach (outlined in Fig.~\ref{fig:figure 1}) is able to unify many advantages of previous metrics and not only detects but also \textit{quantifies} these phenomena, allowing us to locate pockets of parameter space that display maximal ergodicity \cite{Rodriguez-Nieva_2024} and 
enhanced sensitivity of eigenstates to local perturbations \cite{Kim_2025}.
Furthermore, the locality of our metrics
allows us to utilize modern techniques for efficient parallel measurement schemes such as classical shadow tomography \cite{Aaronson_2018,Huang_2020,Elben_2022}.
Our protocol can be executed with approximate eigenstates with non-zero energy variance, reducing the burden of preparing exact eigenstates in experiments~\cite{Yang_2020,Vasilyev_2020,Irmejs_2024,Garratt_2024}.
These findings establish our metrics as powerful tools in the understanding of quantum chaos and ergodicity that also elucidate their connection to the field of quantum learning.

Furthermore, we find that our approach
has deep connections to existing work on using adiabatic deformations as a probe for quantum chaos \cite{Pandey_2020,LeBlond_2021,Kim_2025}.
These works use the norm of the operator known as the adiabatic gauge potential (AGP) to quantify the sensitivity of the spectrum to a specific adiabatic deformation as a probe for chaos.
However, by restricting the locality of the AGP and considering all such adiabatic deformations, one can map this metric directly onto the output of the learning algorithm in Ref.~\cite{Qi_2019}.
Our metrics therefore have natural ties to existing chaos metrics and can be generalized to make contact with previous results in the field.
% }

The rest of the paper is structured as follows. 
In Section~\ref{sec:robustness of H learning}, we review the Hamiltonian learning algorithm from Ref.~\cite{Qi_2019} that
forms the basis of 
our metrics and numerically demonstrate that ergodicity increases the robustness of the learning algorithm.
We then analytically show that ergodicity not only implies the robustness of the algorithm but also that a large class of integrable systems is parametrically less robust to errors during Hamiltonian learning.
In Section~\ref{sec:general learning}, we generalize our arguments to arbitrary algorithms that learn a Hamiltonian from a single one of its eigenstates and discuss the connection between the AGP norm and the learning algorithm from Ref.~\cite{Qi_2019}.
% %
In Section~\ref{sec:quantifying ergodicity}, we show that our metrics go beyond simply identifying ergodicity by continuously quantifying it, allowing us to locate pockets of ``maximal ergodicity'' in parameter space first found in Ref.~\cite{Rodriguez-Nieva_2024}.
% %
In Section~\ref{sec:probing chaos}, we show how our metrics also quantify other aspects of chaos by detecting regions of parameter space where the susceptibility of eigenstates to small local perturbations is maximized \cite{Kim_2025}.
% %
Finally, in Section~\ref{sec:state prep and measurement}, we discuss the experimental applicability of our metrics using realistic approximate eigenstates and randomized measurements.

\section{Robustness of Hamiltonian Learning}
\label{sec:robustness of H learning}

We begin by examining the robustness of Hamiltonian learning methods to noise, focusing on the protocol introduced in Ref.~\cite{Qi_2019} that we use to define our metrics.
First, we will numerically analyze its behavior in various integrable and ergodic spin systems.
We will then use analytic arguments to generalize our results to a wide range of both spin systems and learning algorithms.

The algorithm from Ref.~\cite{Qi_2019} learns the Hamiltonian of a multi-qubit system from a single energy eigenstate by minimizing the variance of an operator ansatz for the Hamiltonian on that state.
Specifically, we assume the Hamiltonian has the form 
\begin{align}
    H = \sum_{\alpha=0}^{N_L} c_{\alpha} L_{\alpha} ,
\end{align}
where $\{L_{\alpha}\}$ is a basis of $N_L$ operators that are all $\ell$-local and $c_{\alpha}$ are coefficients.
\edit{
In this work, we will differentiate between $\ell$-local operators, which are built of terms with support on no more than $\ell$ sites, and \textit{geometrically} $\ell$-local operators, where those $\ell$ sites must be adjacent to each other.
Throughout}, we will consider the basis of all geometrically 2-local Pauli strings (i.e., operators that have nontrivial support on no more than two neighboring sites).
To learn an unknown Hamiltonian $H$ of this form, we are given a single eigenstate $\ket{v}$.
The algorithm proceeds by constructing the covariance matrix
\begin{align}
    M_{\alpha \beta} = \frac{1}{2} \bra{v} \{ L_{\alpha}, L_{\beta} \} \ket{v} - \bra{v} L_{\alpha} \ket{v} \bra{v} L_{\beta} \ket{v} ,
    \label{eq:cov matrix}
\end{align}
where $\{,\}$ is the anticommutator.
Because $M$ is real and symmetric, we can diagonalize it to find real eigenvalues $\sigma_a^2$ and eigenvectors $o^{(a)}_{\alpha}$.
These eigenvectors correspond to operators $\mathcal{O}_a = \sum_{\alpha} o^{(a)}_{\alpha} L_{\alpha}$ with variance $\sigma_a^2$, since
\begin{align}
    \bra{v} \mathcal{O}_a^2 \ket{v} - \bra{v} \mathcal{O}_a \ket{v}^2 = \sum_{\alpha \beta} o^{(a)}_{\alpha} M_{\alpha \beta} o^{(a)}_{\beta} = \sigma_a^2 .
    \label{eq:var spectrum}
\end{align}
We therefore refer to the values $\sigma_a^2$ as the $\ell$-local \textit{variance spectrum} of state $\ket{v}$.
\edit{
Note that although the operators in Eq.~\eqref{eq:cov matrix} need not be geometrically local, they are still $2\ell$-local.
}

Clearly, the Hamiltonian will correspond to a vector in the kernel of $M$, since by definition the variance of $H$ on state $\ket{v}$ is zero.
Thus, if there is a single zero eigenvalue of $M$, the Hamiltonian is uniquely constructed from $\ket{v}$ up to an overall scale factor.
It has been shown that $M$ almost always has a unique zero eigenvalue \cite{Qi_2019}.
\edit{Specifically, Ref.~\cite{Qi_2019} assumes a fixed Hilbert space and local operator basis and the guarantee that at least one Hamiltonian for this system has a unique zero eigenvalue of $M$ for the $i$th eigenstate.
These assumptions are sufficient to prove that the set of Hamiltonians where this eigenvalue is not unique 
% (for the $i$th eigenstate) 
has measure zero; in other words, the set of Hamiltonians that cannot be learned from the $i$th eigenstate has measure zero.}
Importantly, however, a separate issue can still plague Hamiltonian reconstruction: noise.
Constructing $M$ from $\ket{v}$ will generally suffer from some kind of error, whether it be from projective measurements, decoherence, or imperfect state preparation.
Such an effect can be modeled as a perturbation $\Delta M$ to the exact $M_0$ such that the constructed matrix is $M = M_0 + \epsilon \Delta M$.
This error will then propagate to the learned Hamiltonian as an error in the zero eigenvector of $M$: $H = H_0 + \Delta H$.
In fact, there could be no operator in the kernel of $M$, in which case we take the operator associated with the smallest eigenvalue of $M$ as an alternate, approximate solution.
By standard perturbation theory arguments \cite{Qi_2019,Lindner-Bairey_2019} reviewed in Section~\ref{sec:analytic results}, the magnitude of the Hamiltonian error $\Delta H$ for a fixed magnitude of covariance error $\epsilon \Delta M$ is determined by the \textit{spectral gap} of $M_0$, given by $\sigma_1^2 - \sigma_0^2$.
The larger this gap is, the smaller the magnitude of $\Delta H$, and vice versa.
As such, the gap of the covariance matrix directly quantifies the robustness of Hamiltonian learning\edit{, as originally discussed in Ref.~\cite{Qi_2019}}.

\subsection{Numerical Investigation}
\label{sec:numerical investigation}

\begin{figure*}[t]
    \centering
    \includegraphics[width=\linewidth]{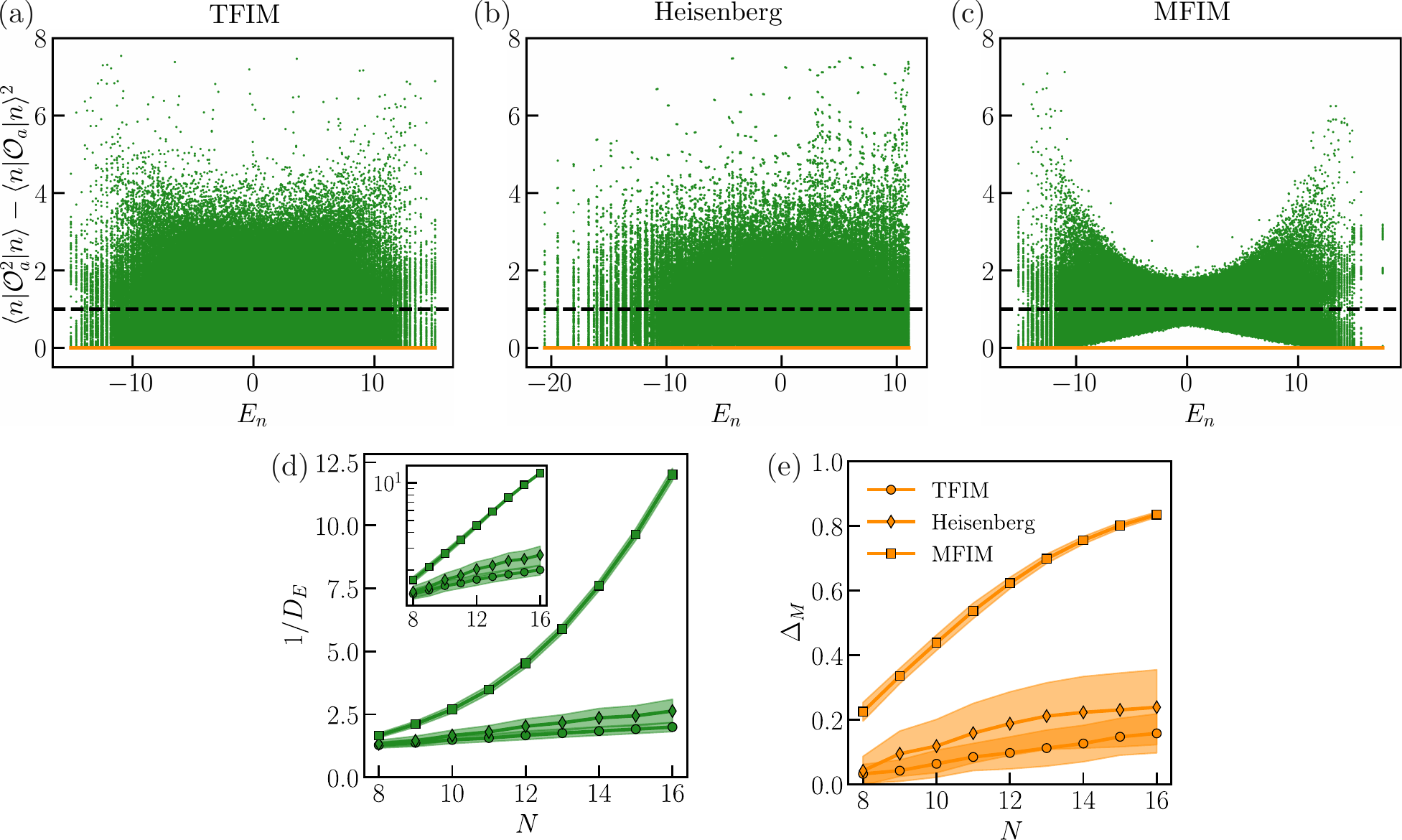}
    \caption{
    \textbf{Variance Spectra in Integrable and Ergodic Hamiltonians.}
    We calculate the variance spectra for all eigenstates of the transverse field Ising model in (a) [Eq.~\eqref{eq:Ising Hamiltonian}, $g = 1$, $h = 0$], the Heisenberg model in (b) [Eq.~\eqref{eq:XXZ Hamiltonian}, $\Delta = 1$], and the mixed field Ising model in (c) [Eq.~\eqref{eq:Ising Hamiltonian}, $g = 1$, $h = 0.3$] at system size $N=12$. 
    The Hamiltonian is always learned as a zero-variance operator (orange), whereas the rest of the spectra (green) show a clear qualitative difference between integrable and ergodic models. 
    We calculate the inverse spread $1/D_E$ [Eq.~\eqref{eq:D_E definition}] in (d) and the gap $\Delta_M$ [Eq.~\eqref{eq:Delta_M definition}] in (e) of the spectra for each model averaged over a microcanonical distribution of eigenstates at energy $E = 0$ (the ribbons denote the standard deviation of each metric over the ensemble).
    The inset in (d) plots $1/D_E$ on a log scale to show exponential growth.
    The inverse spread grows parametrically faster with system size $N$ in the ergodic model than in the integrable ones, and the variance gap similarly increases more quickly. We derive this parametric difference between the ergodic and free fermion (TFIM) models in Appendices \ref{app:rmt predictions} and \ref{app:ff predictions}.
    }
    \label{fig:vars vs energy}
\end{figure*}

We will now analyze the robustness of this learning algorithm in various integrable and ergodic spin-1/2 chains to show that it can be used to quantify ergodicity.
In particular, we show in Fig.~\ref{fig:vars vs energy} that ergodic and integrable systems have qualitatively different variance spectra such that ergodic systems allow for drastically more robust Hamiltonian learning.
These results provide intuition and lay the groundwork for our analytic results in Section~\ref{sec:analytic results} and \ref{sec:general learning}, where we demonstrate how ergodicity implies parametrically more robust learning across many different systems and learning algorithms.

The three spin-1/2 chains we consider are the transverse field Ising model (TFIM) as an example of a free fermion system, the XXZ model at the Heisenberg point as an example of an interacting integrable system, and the mixed field Ising model (MFIM) as an example of an ergodic system.
The first and third systems are defined by the Hamiltonian from \cite{Rodriguez-Nieva_2024}:
\begin{align}
    H_{\rm IM} = h_1 Z_1 + h_N Z_N + \sum_{i=1}^{N-1} Z_i Z_{i+1} + \sum_{i=1}^{N} g X_i + h Z_i ,
    \label{eq:Ising Hamiltonian}
\end{align}
where $X_i,Y_i,Z_i$ are the spin-$1/2$ Pauli operators on the $i$th site of a 1D chain with open boundary conditions such that translation invariance is broken.
The parameter $g$ ($h$) determines the transverse (longitudinal) field strength, and the additional boundary fields are fixed at $h_1 = 0.25$ and $h_N = -0.25$ to break inversion symmetry.\footnote{Breaking these explicit symmetries allows us to compute spectral metrics like the level spacing ratio without having to first restrict to a symmetry sector.
\edit{
For a more in-depth discussion of the subtleties associated with the choice of boundary conditions, see Appendix~\ref{app:boundary conditions}.
}
}
When $g=0$ or $g \to \infty$, the model is classically integrable,
whereas $h=0$ corresponds to free fermions via the Jordan-Wigner transformation \cite{Schultz_1964,Dziarmaga_2005}.
For our TFIM calculations, we use $g = 1$ in this transverse field limit.
Away from these three limits, the model is nonintegrable.
In this mixed field regime, we will default to choosing $g = 1$ and $h=0.3$ \cite{Rodriguez-Nieva_2024}.

The XXZ model we study is described by the Hamiltonian
\begin{align}
    H_{\mathrm{XXZ}} = h_1 Z_1 + \sum_{i=1}^{N-1} X_i X_{i+1} + Y_i Y_{i+1} + \Delta Z_i Z_{i+1} ,
    \label{eq:XXZ Hamiltonian}
\end{align}
where $\Delta$ is the anisotropy parameter and $h_1 = 0.05$ is fixed at a small value to break the model's inversion symmetry \cite{Brenes_2020} without destroying integrability \cite{Alcaraz_1987,Gubin_2012,Kim_2024}.
Under the Jordan-Wigner transformation, this model maps to interacting fermions 
that can be solved exactly using the Bethe ansatz \cite{Orbach_1958,Yang_1966,Korepin_Bogoliubov_Izergin_1993} and as such avoids thermalization.
Furthermore, this model possesses a $U(1)$ symmetry generated by the total $z$ magnetization $m = \sum_i Z_i$, and as such we will restrict our simulations to the largest magnetization sector ($m = 0$ for $N$ even and $m=1$ for $N$ odd) except in Fig.~\ref{fig:vars vs energy}(b) where we plot the entire energy spectrum.
Finally, we will set $\Delta = 1$ from now on, restricting this model to the Heisenberg point.
% \edit{\sout{with an $SU(2)$ symmetry}}.

Given these models, let us now diagonalize each Hamiltonian and find the variance spectrum of each eigenstate.
As mentioned above, we take as our operator basis the set of all geometrically 2-local Pauli strings.
For each eigenstate $\ket{n}$ of a given Hamiltonian, we compute the covariance matrix from Eq.~\eqref{eq:cov matrix}, diagonalize it, and plot the resulting spectrum as a function of the eigenstate energy $E_n$ in Fig.~\ref{fig:vars vs energy}(a-c).
We plot the first eigenvalue $\sigma_0^2$ (which corresponds to $H$ and therefore vanishes) in orange and the rest of the spectrum in green.
By comparing the TFIM [Fig.~\ref{fig:vars vs energy}(a)] and the Heisenberg model [Fig.~\ref{fig:vars vs energy}(b)] to the MFIM [Fig.~\ref{fig:vars vs energy}(c)], it is immediately clear that the emergence of ergodicity has a dramatic effect on the spread of the variance spectrum, especially near the midspectrum corresponding to infinite effective temperature.
Indeed, for the integrable models, eigenstates have variance spectra that generally spread from small values near 0 up beyond 1 across the entire energy spectrum.
In stark contrast, the variance spectra for the MFIM become highly concentrated around 1 (aside from $\sigma_0^2$) as the eigenstates approach infinite temperature and ergodicity emerges.

We have now seen that Hamiltonian learning is more robust for ergodic systems. However, the question remains if this is sufficient to accurately distinguish integrable from ergodic. 
To this end, we consider two metrics.
The first is simply the gap of the variance spectrum as discussed above, defined more generally as
\begin{align}
    \Delta_M = \sigma_{|K|}^2 - \sigma_{0}^2,
    \label{eq:Delta_M definition}
\end{align}
where $K$ denotes the set of values $a$ for which $\sigma_a^2 = 0$.
For example, in the Ising model, $K = \{0\}$ since the Hamiltonian is the only operator in the kernel of $M$, whereas the XXZ model has $K = \{0,1\}$ since both the Hamiltonian and the magnetization $m$ have zero variance.
Therefore, $\sigma_{|K|}^2$ denotes the smallest nonzero eigenvalue of $M$.
As such, this definition accounts for the fact that $\sigma_1^2 - \sigma_0^2 $ always vanishes in the XXZ model and instead uses $\sigma_2^2 - \sigma_0^2$ to quantify the robustness of the learning algorithm. 
\edit{
Also note that this definition of $\Delta_M$ still holds even if $M$ has no zero modes, which is the case if we e.g. do not use an exact eigenstate to construct $M$ (see Section~\ref{sec:state prep and measurement}).
}
The second metric more directly quantifies the spread of the spectrum:
\begin{align}
    D_E^2 = \frac{1}{N_L - |K|} \sum_{a\notin K} (\sigma_a^2 - 1)^2.
    \label{eq:D_E definition}
\end{align}
Intuitively, this metric simply quantifies how much the nonzero values of $\sigma_a^2$ spread around 1.
In Section~\ref{sec:analytic results}, we will show that this reference value of 1 is predicted by ETH at infinite temperature, and we will furthermore show that $D_E$ quantifies how much the robustness of the learned Hamiltonian depends on the particular error that occurs.
More specifically, a small $D_E$ implies that the algorithm is uniformly robust against all errors in $M$, whereas a large $D_E$ implies that the algorithm is robust against some errors and sensitive to others.

We plot $1/D_E$ and $\Delta_M$ (such that both increase for more ergodic systems) for each of our three models in Fig.~\ref{fig:vars vs energy}(d-e) as a function of system size after averaging over microcanonical ensembles of eigenstates centered around energy $E = 0$.
Following Ref.~\cite{Rodriguez-Nieva_2024}, we choose ensemble sizes of 100, 150, 200, 300, 400, 500, 600, 1000, and 2000 for system sizes $N=8$ through 16.
The MFIM shows an exponential growth of $1/D_E$ (seen in the inset)
in stark contrast to the slower polynomial growth in the TFIM and Heisenberg models \cite{Pandey_2020,Alba_2015,Essler_2024,LeBlond_2019}.
We use the ribbons to display the standard deviation over the ensemble (rather than the standard error of the mean) to show how much each metric fluctuates across different eigenstates.
Although these fluctuations are smaller in the ergodic regime as expected, they are still small enough in the integrable models
to suggest that a single midspectrum eigenstate is enough to accurately distinguish integrable and ergodic systems by calculating the variance spectrum.

Similarly, the variance gap $\Delta_M$ approaches $1$ much more quickly in the MFIM than in the two integrable models.
Although the fluctuations about the means are much larger in the integrable models for this metric, there is still a clear quantitative difference between the integrable and ergodic limits, and this difference is amplified as $N$ grows.
As such, we directly confirm that Hamiltonian learning is not only more robust in the ergodic regime but moreover that this difference is large enough to accurately distinguish integrable and ergodic systems.

\subsection{Analytic Results}
\label{sec:analytic results}

While these numerical results show a clear relationship between ergodicity and the robustness of Hamiltonian learning in specific systems [Eqs.~\eqref{eq:Ising Hamiltonian} and \eqref{eq:XXZ Hamiltonian}], we now analytically show that (1) ergodicity in general implies the robustness of Hamiltonian learning and (2) a large class of integrable models (in particular, translation-invariant free fermion models) are parametrically less robust.
These results, in conjunction with the numerics above, establish the robustness of Hamiltonian learning as a general metric for ergodicity.

\textit{Quantifying Sensitivity.}
We begin by precisely quantifying the susceptibility of the learned Hamiltonian to errors in $M$\edit{, following the analysis in Ref.~\cite{Qi_2019}}.
As mentioned above, we express the effect of a general error in the covariance matrix as
\begin{align}
    M = M_0 + \epsilon \Delta M .
    \label{eq:perturbed M}
\end{align}
For simplicity, we will assume that the Hamiltonian is the unique operator in the kernel of $M_0$ (i.e. $|K| = 1$).
We can now ask how the estimated Hamiltonian, which is encoded as the ground ``operator'' of $M_0$, changes under this perturbation.
Let us express the eigenoperators $\mathcal{O}_a$ as elements $|\mathcal{O}_a)$ of the operator vector space with inner product $(A|B) = \Tr(A^{\dagger} B)/d$ (where $d = 2^N$ is the Hilbert space dimension).
Then we can express the change in the learned Hamiltonian from $\Delta M$ as $|H) = |H_0) + |\Delta H)$.
For infinitesimal $\epsilon$, we can write $|\partial_{\epsilon}H) = |\Delta H)/\epsilon$, and first-order perturbation theory tells us that
\begin{align}
    |\partial_{\epsilon} H) = \sum_{a>0} \frac{(\mathcal{O}_a|\Delta M|H_0)}{\sigma_0^2 - \sigma_a^2} |\mathcal{O}_a)  ,
\end{align}
where $\sigma_a^2$ are the eigenvalues of $M_0$.
The magnitude of this vector then encodes the Hamiltonian's susceptibility $\chi_H$ to the change in $M$, since
\begin{align}
    \chi_H = (\partial_{\epsilon} H|\partial_{\epsilon} H) - |( H_0|\partial_{\epsilon} H)|^2
    \label{eq:chi H}
\end{align}
defines the susceptibility and $( H_0|\partial_{\epsilon} H) = 0$.

Let us now understand how this susceptibility changes for different choices of perturbation $\Delta M$.
\edit{If we consider $\Delta M$ with a fixed magnitude $\mu$, $\chi_H$ is maximized for $(\mathcal{O}_a|\Delta M|H_0) = \mu \delta_{a,1}$ such that $\Delta M$ only hybridizes $H_0$ with the next operator in the variance spectrum.
%
% If we fix $(\mathcal{O}_1|\Delta M|H_0)=1$, we find
Thus,
\begin{align}
    \sqrt{\chi_H} = \frac{\mu}{\sigma_1^2}, 
\end{align}
}which precisely relates the sensitivity of the learning algorithm to the minimum variance in the spectrum of operators (besides the Hamiltonian) as stated \edit{above and in Ref.~\cite{Qi_2019}}.
Similarly, $\chi_H$ is minimized if we choose \edit{$(\mathcal{O}_a|\Delta M|H_0) = \mu \delta_{a,N_L}$ such that $\sqrt{\chi_H} = \mu/\sigma_{N_L}^2$}, which encodes the \textit{maximum} precision of the learned Hamiltonian if we assume the perturbation only hybridizes the eigenoperators at the edges of the variance spectrum.\footnote{\edit{We note that, in contrast to $\sigma_1^2 \leq N_L/(N_L - 1)$, $\sigma_{N_L}^2$ is only bounded as $O(N_L)$. These both follow from the observation that $\Tr M \leq N_L$.}}
Finally, one can ask how much this susceptibility fluctuates for different choices of $\Delta M$ by calculating the spread of the variance spectrum.
We have already seen how the fluctuations and maximum of $\chi_H$ (captured by the metrics $D_E$ and $\Delta_M$ respectively) act as accurate indicators of ergodicity, and in Section~\ref{sec:probing chaos} we will show that the \textit{minimum} of $\chi_H$ can probe chaos by quantifying the sensitivity of eigenstates to $\ell$-local perturbations.

\textit{Ergodic Systems.}
Now, let us analytically confirm that ergodicity implies the robustness of Hamiltonian learning by ensuring $\chi_H$ is small for all possible errors $\Delta M$.
\edit{
We note that the effect of ETH on this learning algorithm was also explored in Ref.~\cite{Qi_2019}.
To understand how ergodicity affects the covariance spectrum, we briefly review ETH and its predictions, mainly following the treatment in \cite{D_Alessio_2016}.
The eigenstate thermalization hypothesis states that the matrix elements of a low-weight operator $O$ in the energy eigenbasis follow \cite{Deutsch_1991,Srednicki_1994,D_Alessio_2016,Deutsch_2018}
\begin{align}
    \bra{m} O \ket{n} = O(\Bar{E}) \delta_{mn} + e^{-S(\Bar{E})/2} f_O(\Bar{E},\omega) R_{mn},
\end{align}
where $E_m$ ($E_n$) is the energy of eigenstate $m$ ($n$), $\Bar{E} = (E_m + E_n)/2$, $\omega = E_n - E_m$, and $S(\Bar{E})$ is the thermodynamic entropy at energy $\Bar{E}$.
$O(\Bar{E})$ and $f_O(\Bar{E},\omega)$ are both smooth functions of their arguments, and the former is equal to the expectation value of $O$ in a microcanonical ensemble at energy $\Bar{E}$ \cite{D_Alessio_2016}.
$R_{mn}$ is a zero mean, unit variance random variable accounting for fluctuations between different matrix elements, and the symmetry class of this variable (e.g. time reversal invariance) is the same as that of the Hamiltonian.
Because $S$ is generally extensive in system size, ETH implies that expectation values of low-weight operators follow the predictions of the microcanonical ensemble (or generic one in the thermodynamic limit) up to exponentially small fluctuations.
Note that the predictions of ETH have even been numerically verified even for low-weight observables that are not geometrically local \cite{D_Alessio_2016}.
As a result, energy eigenstates look thermal when restricted to a small subsystem: the reduced density matrix on the subsystem approaches \cite{D_Alessio_2016,Garrison_2018}
\begin{align}
    \rho_A = \Tr_{\Bar{A}} \ket{n} \bra{n} \approx \Tr_{\Bar{A}} \frac{e^{-\beta(E_n) H}}{Z},
\end{align}
where $Z$ is the partition function and $\beta(E_n)$ is the temperature that satisfies $\Tr [H e^{-\beta(E_n)H}/Z] = E_n$.
}

\edit{
Notably, these predictions differ from the simple predictions of RMT due to the varying factors $O(\Bar{E})$ and $f_O(\Bar{E},\omega)$, which depend on both the energy mean and difference of the states in question.
However, if we restrict our attention to a small enough energy window $\omega$, these functions can be approximated by constants, and RMT gives accurate predictions for observable matrix elements.
The energy scale below which RMT is accurate and $f_O$ looks constant is referred to as the Thouless energy $E_{\rm Th}$ \cite{D_Alessio_2016,Gharibyan_2018} and has a corresponding Thouless time $t_{\rm Th} \sim 1 / E_{\rm Th}$ after which RMT correlations emerge.
While the Thouless time does grow with system size \cite{D_Alessio_2016,Schiulaz_2019}, it is generally exponentially shorter than the Heisenberg time, defined as $t_{\rm H} \sim 1/ E_{\rm H}$ where the Heisenberg energy $E_{\rm H}$ is defined as the average level spacing \cite{Schiulaz_2019,Pandey_2020,D_Alessio_2016}.
% (states with energy fluctuations below this value are effectively energy eigenstates). 
%
In other words, the predictions of RMT apply in a wide energy regime $E_{\rm H} \lesssim \omega \lesssim E_{\rm Th}$.
}

\edit{
Given these considerations, 
any midspectrum eigenstate of an ergodic Hamiltonian (that follows ETH)
should look random aside from the requirement that the Hamiltonian and any other conserved quantity have zero variance. 
More specifically, we assume that eigenstates in the middle of the spectrum (at infinite temperature such that $\beta = 0$)}
can be modeled as Haar random when considering observables that have no overlap with conserved operators, and we will see that this assumption agrees well with our numerics.
First, we ask what the expected value of any variance $\sigma_a^2$ for $a \notin K$ is when averaging over a Haar random distribution of eigenstates.
This is given by
\begin{align}
    \mathbb{E} [\sigma_a^2] = 1 - \frac{1}{d+1},
\end{align}
where $\mathbb{E}$ denotes the expected value over the Haar distribution of states.
This result agrees well with the concentration of variance values in Fig.~\ref{fig:vars vs energy}(c) around 1.
To make contact with Eq.~\eqref{eq:D_E definition}, we then ask what the expected value is for the deviation of any such $\sigma_a^2$ from 1.
\edit{This is given by
\begin{align}
    \mathbb{E} [(\sigma_a^2 - 1)^2] \leq  \frac{N_L - 1}{d} + O \left( \frac{1}{d^2} \right) .
    \label{eq:ergodic variance}
\end{align}
}The derivation of these values can be found in Appendix \ref{app:rmt predictions}.\footnote{An important caveat to these calculations is that an operator can be orthogonal to the Hamiltonian but have nonzero overlap with $H^2, H^3,\dots$, thus shifting $\sigma_a^2$ away from the ETH prediction.
We address this concern in Appendix~\ref{app:krylov overlaps}, where we verify that our numerics are not affected at the chosen system sizes.}
These results imply that a midspectrum eigenstate of an ergodic Hamiltonian will have a variance spectrum consisting of the Hamiltonian and \edit{geometrically} $\ell$-local conserved quantities with $\sigma_a^2 = 0$ and all other operator variances tightly clustered around $\sigma_a^2 = 1$.
Furthermore, this clustering becomes tighter as the system size grows in the sense that 
$D_E$
decreases at least as fast as $\mathrm{poly}(N)/\mathrm{exp}(N)$, in agreement with the exponential growth of $1/D_E$ in Fig.~\ref{fig:vars vs energy}(d) (see inset).

\begin{figure}[t!]
    \centering
    \includegraphics[width=0.9\linewidth]{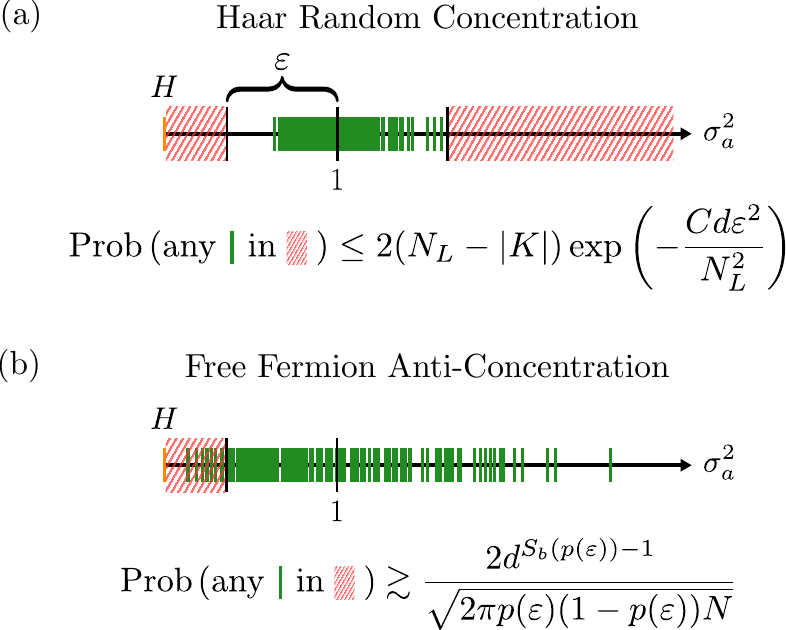}
    \caption{
    \edit{
    \textbf{Concentration and Anti-Concentration of Variance Spectra.}
    }
    (a) By modeling ergodic midspectrum eigenstates as Haar random, we derive an \textit{upper} bound [Eq.~\eqref{eq:levy bound for var spectrum}] on the probability that any nonzero variance $\sigma_a^2 \neq 0$ is more than $\varepsilon$ away from 1.
    This probability decays like an exponential in the Hilbert space dimension $d$, demonstrating the robustness of Hamiltonian learning in the ergodic regime.
    (b) For translationally invariant free fermion systems, we derive an (approximate) \textit{lower} bound [Eq.~\eqref{eq:ff anti-concentration}], consistent in the limit of large $N$, on the probability that any nonzero variance is smaller than $1-\varepsilon$ for an eigenstate selected uniformly at random.
    This probability decays like a power law in $d$, showing that Hamiltonian learning is parametrically less robust in this class of integrable systems than in ergodic systems.
    This bound is not constructed from conserved quantities; rather, we multiply the Hamiltonian by a sinusoidal spatial profile to construct an operator with low variance for an exponential number of eigenstates [see Appendix~\ref{app:ff predictions}].
    }
    \label{fig:analytics cartoon}
\end{figure}

In fact, we can make an even \textit{stronger} statement about the clustering of the variance spectrum using concentration bounds \cite{Mele_2024,Ledoux_2001}.
In particular, using our assumption that $\ket{v}$ can be treated as a Haar random state, the probability that any \edit{geometrically} $\ell$-local operator's variance (besides our conserved quantities) differs from 1 by more than $\varepsilon$ is bounded from above by
\edit{
\begin{align}
    \mathrm{Prob}\left( \bigcup_{a \notin K} \left| \sigma_a^2 - 1 \right| \geq \varepsilon \right) \leq 2 (N_L - |K|) \exp \left( - \frac{C d \varepsilon^2}{ N_L^2 } \right) ,
    \label{eq:levy bound for var spectrum}
\end{align}
}where $C$ is some absolute constant.
This bound is shown graphically in Fig.~\ref{fig:analytics cartoon}(a) and is derived in Appendix \ref{app:rmt predictions}.
Up to polynomial corrections from the $N_L$ factors, this probability decreases as a \textit{double} exponential in system size, drastically limiting the chance that $\Delta_M \ll 1$ as $N$ increases.
For the correct choice of constant $C$, this bound is consistent with the growth of $\Delta_M$ seen in Fig.~\ref{fig:vars vs energy}(e).
Taken together, these results imply that $1/D_E$ and $\Delta_M$ grow rapidly with $N$ in ergodic systems, ensuring that Hamiltonian learning is robust against all errors.

\textit{Free Fermion Systems.}
This concentration of the variance spectrum follows from modeling ergodic eigenstates as Haar random, and as such, we expect large deviations from this clustering in integrable systems where this assumption breaks down.
Indeed, fluctuations around high temperature expectation values in integrable systems tend to decay only polynomially in system size \cite{Alba_2015,Essler_2024,LeBlond_2019}, showing a clear distinction between ergodic and integrable systems, corroborated by the data in Fig.~\ref{fig:vars vs energy}(d).
However, we can go further by making analytic statements about the anti-concentration of the variance spectrum to contrast with Eq.~\eqref{eq:levy bound for var spectrum}.
In Appendix~\ref{app:ff predictions}, we show that the simple solutions of translation-invariant free fermion models allow us to construct explicit operators that are (1) orthogonal to the Hamiltonian and (2) have a small variance for an exponential number of eigenstates in the thermodynamic limit.
As such, we can make an approximate lower bound (using approximations consistent in the limit of large $N$) on the probability that any variance in the spectrum of an average eigenstate is less than $1 - \varepsilon$ as
\begin{align}
    \mathrm{Prob}\left( \bigcup_{a \notin K} \sigma_a^2 \leq 1 - \varepsilon \right) 
    \gtrsim 
    \frac{2 d^{S_b(p(\varepsilon)) - 1}}{\sqrt{2 \pi p(\varepsilon) (1-p(\varepsilon)) N}} .
    \label{eq:ff anti-concentration}
\end{align}
This bound is shown graphically in Fig.~\ref{fig:analytics cartoon}(b).
Here, $S_b(p)$ is the binary entropy function and $p(\varepsilon)$ is a linear function of $\varepsilon$ [defined in Eq.~\eqref{eq:p function definition} in the Appendix] used to count the number of states in the spectrum for which our operator has low variance.
In stark contrast to the ergodic case, this bound shows that translation-invariant free fermion systems have gaps $\Delta_M$ smaller than $1 - \varepsilon$ \textit{at least} as often as the right-hand side of Eq.~\eqref{eq:ff anti-concentration}, approximately upper-bounding the robustness of the learning algorithm as $N \to \infty$.
Furthermore, the numerical results in Fig.~\ref{fig:vars vs energy}(d-e) show that interacting integrable systems behave very similarly to the free fermion case.
As such, ergodicity does not just guarantee the robustness of Hamiltonian learning; its \textit{absence} also implies a \textit{lack} of robustness against certain errors in $M$, establishing these metrics as accurate indicators of ergodicity.

Interestingly, the operators used to construct this inequality do \textit{not} correspond to the conserved quantities $Q$ of the integrable models as one might expect.
In fact, these integrals of motion are generally \textit{less} local than the Hamiltonian in the sense that they have weight greater than $\ell$, and as such they will not be in the span of the operator basis used to construct $M$.
For example, in the TFIM, there are only \textit{two} conserved quantities that are built from geometrically 2-local operators, and one of them is the Hamiltonian.
Beyond these, there are two \edit{geometrically} 3-local integrals, two \edit{geometrically} 4-local integrals, and so on in an infinite hierarchy \cite{Grady_1982,Prosen_1998,fagotti2013reduced,essler2016quench,vidmar2016generalized}, and a similar hierarchy exists in the case of the XXZ model and other Bethe integrable systems \cite{GRABOWSKI_1994,Grabowski_1995}.
Although these integrals of motion will account for some degeneracy in the zero-variance subspace of operators in the variance spectrum, they do not a priori allow for the kind of anti-concentration inequality above.
Furthermore, there does not seem to be a strong reason to expect that truncating these operators to be \edit{geometrically} $\ell$-local would give rise to the observed continuum, as these operators do not possess any sort of tails (as in the case of $l$-bits in MBL \cite{Huse_2014,Dupont_2019,Dupont_2019_2}) that would justify such a truncation.\footnote{Moreover, the eigenstates under consideration are not short-range correlated (even the ground state in Fig.~\ref{fig:vars vs energy}(a) possesses algebraically decaying correlations, as we have tuned the system to the critical point), and as such the small gaps do not immediately follow from the argument found in Ref.~\cite{Qi_2019}.}

Instead, the small gaps present in the TFIM (and other local free fermion models) result from the quasiparticle structure of the spectrum.
More specifically, the fact that fermion modes are either occupied or unoccupied allows us to 
construct low-variance operators by multiplying the Hamiltonian by long-wavelength sinusoidal envelopes reminiscent of arguments used in Lieb-Schultz-Mattis theorems \cite{LIEB_1961,tasaki2022liebschultzmattistheoremtopologicalpoint}.
These envelopes cause the operators to explicitly break translation invariance in a way that marginally changes fermion mode occupations, allowing for a small variance despite being orthogonal to all (translationally invariant) conserved operators $Q$.
For a fuller analytic treatment and numerical demonstration of this construction, see Appendix~\ref{app:ff predictions}.
\edit{
For a discussion of how the operator basis affects this argument, see Appendix~\ref{app:boundary conditions}.
}

To summarize, we have seen that the minimum ($\Delta_M$), maximum ($\sigma^2_{\rm max}$), and inverse spread ($1/D_E$) of nonzero variances $\sigma_a^2$ all quantify the robustness to Hamiltonian learning in different limits, and that $\Delta_M$ and $1/D_E$ both quantitatively distinguish integrable and ergodic systems (we investigate the behavior of $\sigma^2_{\rm max}$ in Section~\ref{sec:probing chaos}).
In particular, $D_E$ decays exponentially in $N$ in ergodic systems and algebraically in integrable systems, showing that the former are robust against all errors while the latter are sensitive to some errors and robust against others.
More concretely, the probability of having $\Delta_M$ smaller than some $1 - \varepsilon$ is \textit{upper} bounded by an exponential decay in $d$ in the ergodic regime but is approximately \textit{lower} bounded by an algebraic decay in $d$ in translationally invariant free fermion systems.
As such, these integrable systems have parametrically smaller variance gaps $\Delta_M$ than ergodic systems and are therefore less robust to errors.

\subsection{Generalization: Information-Theoretic Results via Sensing}
\label{sec:general learning}

Although we have analytically established the relationship between ergodicity and the robustness of Hamiltonian learning, the above results are specific to the algorithm proposed in Ref.~\cite{Qi_2019}.
While this choice of algorithm is well-motivated by its various connections to other algorithms and quantities reviewed in Appendix~\ref{app:variance spectrum connections} \cite{Bentsen_2019,Zhan_2024,Moudgalya_2023_num,Moudgalya_2024,Li_2025,Pawlowski_2025}, we can make even more general statements about how ergodicity affects Hamiltonian learning.
In particular, we now use results from quantum metrology to argue that ergodicity implies the robustness of Hamiltonian learning for any algorithm that learns $H$ from a single eigenstate $\ket{v}$.
This analysis will also elucidate the connection between the variance spectrum and the AGP norm analyzed in \cite{Pandey_2020,LeBlond_2021,Kim_2025}.

\textit{Learning as a Sensing Problem.}
Intuitively, our argument proceeds as follows: in ergodic systems, energy eigenstates are ``sensitive'' in the sense that a small change in the Hamiltonian $H$ causes a large change in the eigenstate $\ket{v}$ \cite{Pandey_2020,LeBlond_2021,Kim_2025}.
If we then consider the inverse problem of deriving $H$ from $\ket{v}$, a small error in the state $\ket{v}$ will translate to an \textit{even smaller} error in the Hamiltonian $H$ by simple propagation of error arguments, showing that the learning algorithm is robust.

To make this line of reasoning quantitative, we begin by defining the learning algorithm as a function $f(\ket{v})$ that maps the eigenstate $\ket{v}$ to its Hamiltonian $H$.
The robustness of the algorithm can then be viewed as a sensing problem: Given two eigenstates $\ket{v(0)}$ and $\ket{v(\lambda)}$ of two respective Hamiltonians $H = f(\ket{v(0)})$ and $H + \lambda V = f(\ket{v(\lambda)})$ that differ by a small shift $\lambda V$, can the learning algorithm distinguish the two states from a given number of measurements by ``sensing'' the perturbation $\lambda V$?
If so, the learning algorithm is robust in the sense that quantum projection noise from a constant number of shots is not enough to conflate the two nearby Hamiltonians.
If not, the shot noise will generate an error large enough to estimate the Hamiltonian $H + \lambda V$ when the true Hamiltonian is $H$.

This interpretation allows us to recast the Hamiltonian learning algorithm as a sensing problem where the parameter to be sensed is $\lambda$.
More specifically, the robustness of the learning algorithm is quantified by the uncertainty in $\lambda$: a smaller value of $\Var(\lambda)$ corresponds to a more robust learning algorithm, since smaller error bars allow us to better distinguish the Hamiltonians $H$ and $H + \lambda V$.
To relate this variance to properties of an ergodic eigenstate $\ket{v} = \ket{v(0)}$,
we use the quantum Cram\'er-Rao bound \cite{Liu_2019}:
\begin{align}
    \Var(\lambda) \geq \frac{1}{F_Q[\mathcal{A}_{\lambda},\ket{v}]} ,
    \label{eq:quantum cramer rao bound}
\end{align}
where 
\begin{align}
    F_Q[\mathcal{A}_{\lambda},\ket{v}] = 4 \left(\bra{v} \mathcal{A}_{\lambda}^2 \ket{v} - \bra{v} \mathcal{A}_{\lambda} \ket{v}^2 \right)
    \label{eq:qfi}
\end{align}
is the quantum Fisher information (QFI) of pure state $\ket{v}$ with respect to perturbation $\lambda$. 
Here, $\mathcal{A}_{\lambda}$ is the AGP \cite{Kolodrubetz_2017} that generates the unitary rotation mapping $\ket{v(0)}$ to $\ket{v(\lambda)}$:
\begin{align}
    \ket{v(\lambda)} = U(\lambda,0)\ket{v(0)} = \mathcal{P} \exp\left[i \int_0^{\lambda} d\lambda' \mathcal{A}_{\lambda}(\lambda')\right] \ket{v(0)} ,
\end{align}
where $\mathcal{P}$ denotes path ordering such that (for $\lambda>0$) operators with larger $\lambda'$ appear to the left of those with smaller $\lambda'$.
In essence, the AGP encodes the evolution of the eigenstate $\ket{v}$ as we tune $\lambda$, and its variance in Eq.~\eqref{eq:qfi} quantifies how much the eigenstate changes for an infinitesimal shift in $\lambda$.
As such, this variance maps onto the fidelity susceptibility, and its average over all eigenstates encodes the AGP's norm as an operator \cite{Pandey_2020,Kim_2025}.

\textit{Connection to Ergodicity.}
It has been established that in ergodic systems, the (appropriately regularized) QFI \eqref{eq:qfi} with respect to local $V$ grows exponentially in system size for an average eigenstate \cite{Pandey_2020,LeBlond_2021,Kim_2025}.
In contrast, free fermion and Bethe integrable systems respectively display constant and polynomial scaling in system size for integrability-preserving perturbations.
The bound in Eq.~\eqref{eq:quantum cramer rao bound} therefore implies that, for such perturbations $\lambda V$, ergodic systems allow for a much more precise estimation of the Hamiltonian from any learning algorithm.
This observation is in agreement with previous work, where ETH is often invoked to argue for the ability to learn Hamiltonians from their eigenstates \cite{Garrison_2018,Qi_2019,Hsieh-Li_2020}.

\begin{figure*}[t!]
    \centering
    \includegraphics[width=\linewidth]{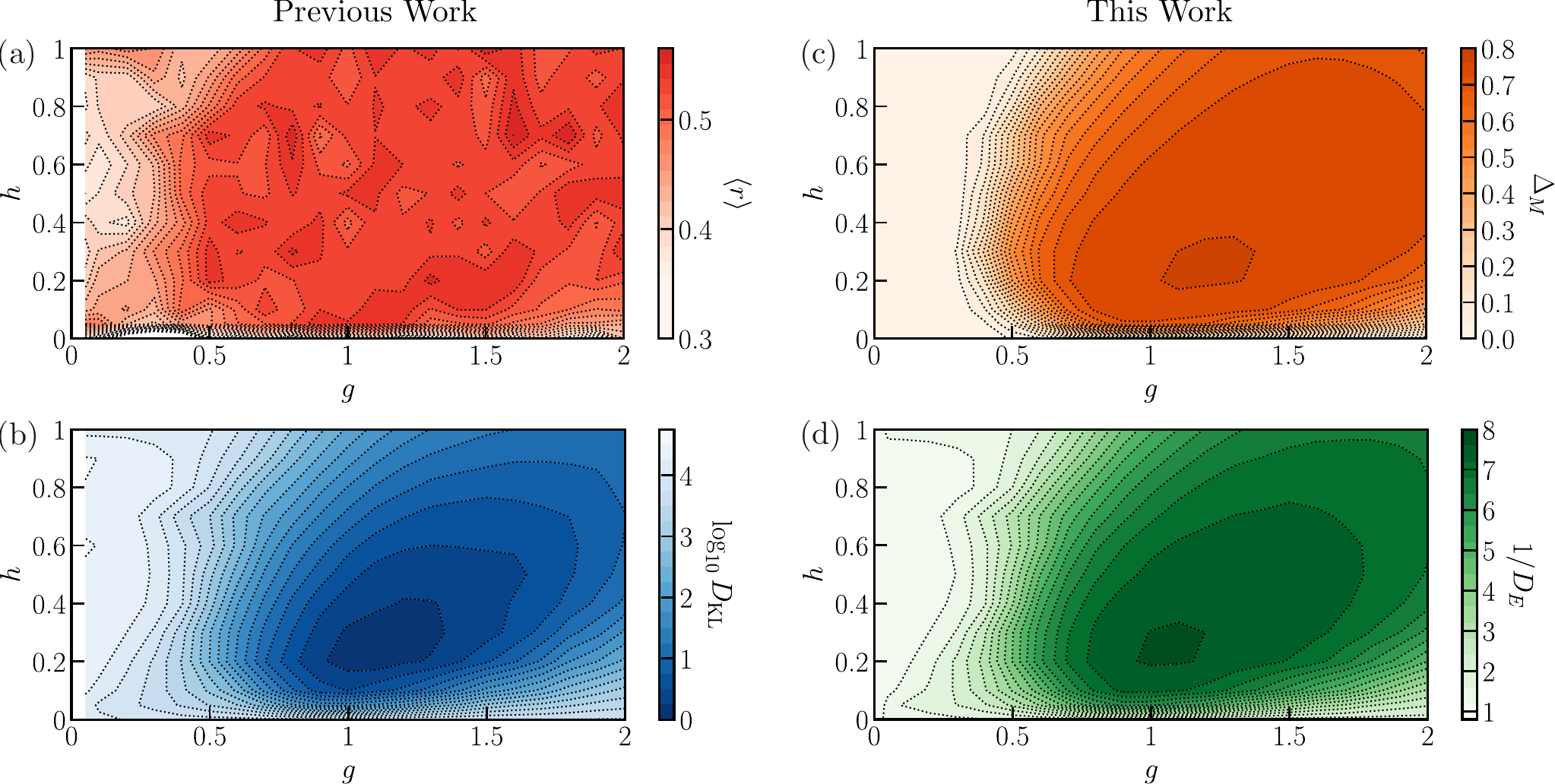}
    \caption{\edit{
    \textbf{Ergodicity Metrics in the Mixed Field Ising Model.}
    }
    Using the $N=14$ mixed field Ising model, 
    we compare established ergodicity metrics to our proposed metrics for a microcanonical distribution of $600$ eigenstates centered at $E=0$.
    (a) The average level spacing $\langle r \rangle$ [Eq.~\eqref{eq:level spacing ratio}] is able to distinguish an ergodic region of parameter space (dark\edit{er} red) from the two integrable lines $g=0$ and $h=0$\edit{, but is plagued by noise for this number of eigenstates}.
    (b) In contrast, the entanglement statistics measure $D_{\rm KL}$ [Eq.~\eqref{eq:KL divergence definition}] from \cite{Rodriguez-Nieva_2024} is not only able to distinguish integrable from ergodic but also reveals structure within the ergodic regime.
    As such, it is able to locate a pocket of maximal ergodicity near $g = 1$, $h = 0.3$, where agreement with the predictions of RMT is maximized.
    We then plot our proposed metrics for ergodicity: the variance gap $\Delta_M$ (c) and inverse spread $1/D_E$ (d)\edit{, averaged over the distribution of eigenstates}.
    Both metrics are able to distinguish between integrable and ergodic regimes and identify the maximally ergodic region 
    originally found with $D_{\rm KL}$.
    The blank regions in (a,b) near $g=0$ come from divergences and numerical instabilities in the metrics for classically integrable systems.
    }
    \label{fig:ergodicity maps}
\end{figure*}

\textit{Relating the Variance Spectrum to the AGP Norm.}
In fact, we can analytically relate the robustness of the learning algorithm from Ref.~\cite{Qi_2019} with the QFI.
Indeed, one can show that the covariance matrix $M$ makes \textit{explicit} use of the QFI to learn the Hamiltonian, since the eigenvalues of $M$ are equivalent to
\begin{align}
    \sigma_a^2 = \frac{1}{4} F_Q[\mathcal{O}_a,\ket{v}].
\end{align}
That is, the $a$th eigenvalue of $M$ equals the QFI of our eigenstate $\ket{v}$ with respect to generator $\mathcal{O}_a$ up to a constant factor $1/4$.
As such, the variance spectrum encodes the QFI of all Hamiltonian perturbations generated by \edit{geometrically} $\ell$-local operators.
Note that these \edit{geometrically} $\ell$-local generators are \textit{not} the same as the (generally nonlocal) AGPs associated with all \edit{geometrically} $\ell$-local perturbations $V$ of the Hamiltonian.
However, one can verify that the $V$ associated with a local generator, given by
\begin{align}
    \lambda V_a = e^{-i\lambda \mathcal{O}_a} H e^{i\lambda \mathcal{O}_a} - H ,
\end{align}
is \edit{geometrically} $(2\ell-1)$-local for infinitesimal $\lambda$ and quasilocal in general due to Lieb-Robinson bounds \cite{lieb1972finite}.
Furthermore, analytic and numerical investigations \cite{Claeys_2019,del_Campo_2012,Bachmann_2017,gjonbalaj2025shortcutsanalogpreparationnonequilibrium,morawetz2025universalcounterdiabaticdriving,finžgar2025counterdiabaticdrivingperformanceguarantees} have shown that truncating an AGP to a local form has the effect of implementing a low-energy ``cutoff'' in the same spirit as the regulator used in \cite{Pandey_2020,LeBlond_2021,Kim_2025} without requiring an explicit choice of the exact cutoff.
As such, the variance spectrum is able to effectively probe the QFI for a specific subset of local Hamiltonian perturbations.

Importantly, this argument for general robustness is qualified by the fact that the QFI fluctuates for different choices of perturbation $\lambda$ and eigenstate $\ket{v}$ \cite{Sierant_2019,Sels_2021}.
As we have seen, the worst-case scenario for estimating $H$ corresponds to the choice of $\lambda$ with the smallest QFI, as this implies a large error in $H$ quantified by $\Var(\lambda)$.
In contrast, the choice $\lambda$ with the largest QFI limits the best-case scenario by bounding the \textit{smallest} corresponding error in $H$.
These two limits can differ dramatically in physical systems:
for example, while perturbations that preserve integrability have QFI that grows at most polynomially in the size of an integrable system,
perturbations that \textit{break} integrability show \textit{exponential} QFI growth 
\cite{Pandey_2020,Brenes_2020_lf,LeBlond_2020,LeBlond_2021}.
Moreover, if one weakly breaks integrability, the QFI does grow exponentially with system size but does so parametrically faster than the ETH prediction.
Indeed, this feature is the foundation for the proposed definition of ``maximally chaotic'' systems that saturate the maximal growth of AGP norm with system size \cite{Kim_2025}.
These features further motivate our use of Ref.~\cite{Qi_2019}'s learning algorithm for our metrics, as the variance spectrum easily quantifies the robustness of learning $H$ to \textit{any} error in $M$.

In summary, by suitably modifying the AGP norm originally discussed in Ref.~\cite{Pandey_2020} to (1) require the AGP to be \edit{geometrically} $\ell$-local for any system size and (2) consider all perturbations $\lambda V$ generated by a \edit{geometrically} $\ell$-local AGP in parallel, we can map this established metric for chaos onto the variance spectrum.
This connection also allows us to argue that learning a local Hamiltonian from a single eigenstate will in general be more robust to errors in ergodic systems, regardless of the algorithm used.
Having established the generality of this relationship, we now turn to the variance spectrum's ability to measure ergodicity and chaos throughout a given model's parameter space.

\section{Quantifying Ergodicity}
\label{sec:quantifying ergodicity}

We now show that our metrics go beyond simply distinguishing integrable and ergodic regimes by continuously quantifying ergodicity \textit{within} the ergodic regime.
This ability to continuously measure ergodicity allows us to construct maps of parameter space that
identify regions of \textit{maximal ergodicity},\footnote{Although Ref.~\cite{Rodriguez-Nieva_2024} refers to these regions as ``maximally chaotic,'' we already use this term to refer to the maximal growth of the AGP norm \cite{Kim_2025}.
Because $D_{\rm KL}$ quantifies agreement with the predictions of ETH, we will instead refer to the regions from Ref.~\cite{Rodriguez-Nieva_2024} as ``maximally ergodic.''}
first found in Ref.~\cite{Rodriguez-Nieva_2024} using the system's half-chain entanglement entropy.

In particular, we take the MFIM of Eq.~\eqref{eq:Ising Hamiltonian} and plot four different metrics as a function of $g$ and $h$.
In Fig.~\ref{fig:ergodicity maps}(a), we consider the level spacing ratio \cite{Oganesyan_2007} defined using 
\begin{align}
    r_n = \frac{\min(\Delta E_n,\Delta E_{n+1})}{\max(\Delta E_n,\Delta E_{n+1})}, 
    \label{eq:level spacing ratio}
\end{align}
where $\Delta E_n = E_n - E_{n-1}$.
After averaging over our microcanonical distribution of $600$ eigenstates centered at $E=0$ to get $\langle r \rangle$, this ratio should approach $0.386$ in integrable systems with Poissonian level statistics and $0.536$ in ergodic systems with Wigner-Dyson statistics \cite{Atas_2013}.

\begin{figure}[t]
    \centering
    \includegraphics[width=\linewidth]{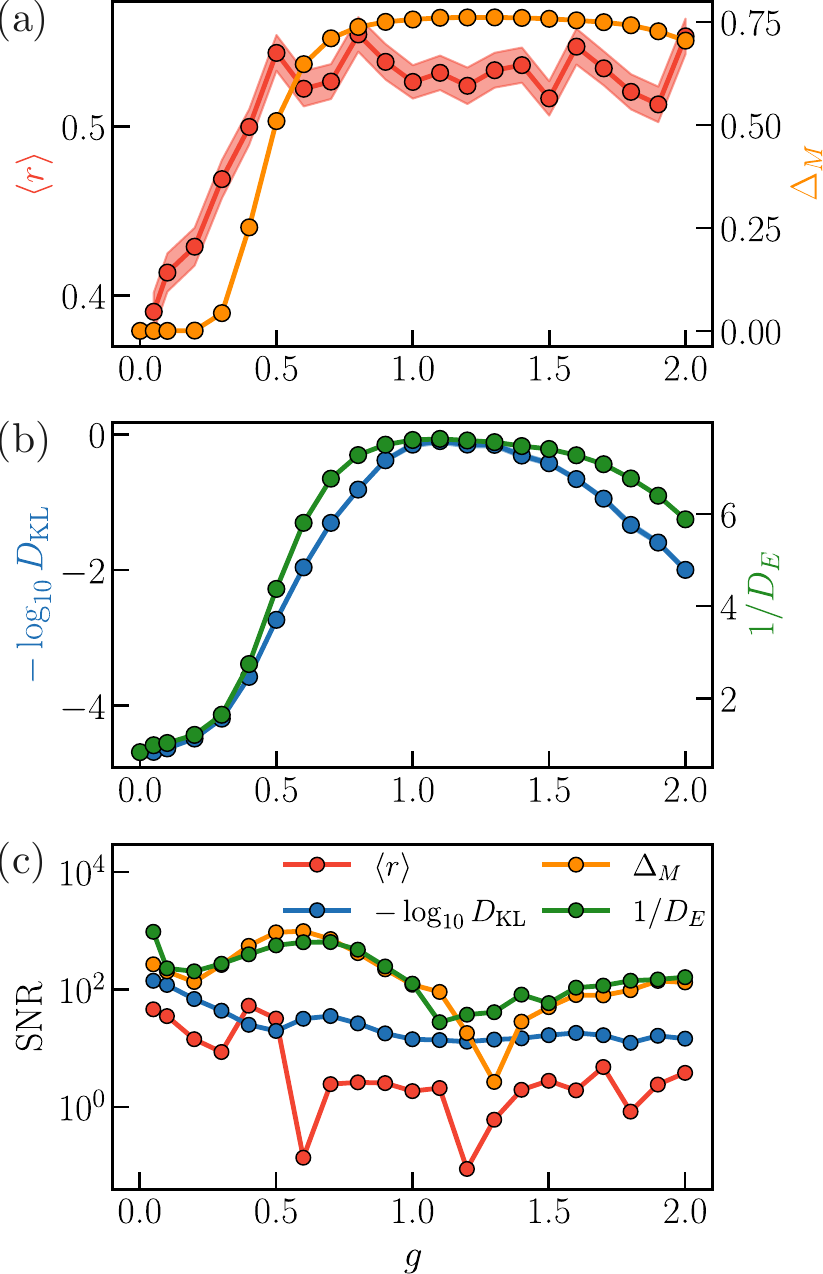}
    \caption{
    \edit{
    \textbf{Sensitivity of Ergodicity Metrics.}
    Using the data from Fig.~\ref{fig:ergodicity maps}, we plot $h = 0.3$ cuts for (a) $\langle r \rangle$ and $\Delta_M$ and (b) $- \log_{10} D_{\rm KL}$ and $1/D_E$.
    We also include estimates of the noise in each metric from sampling over a finite ensemble of eigenstates (standard error of the mean for $\langle r \rangle$, $\Delta_M$, and $1/D_E$, bootstrapped standard error of the mean for $- \log_{10} D_{\rm KL}$) as ribbons; these estimates are essentially invisible for all metrics besides $\langle r \rangle$.
    (c) To approximately quantify the sensitivity of each metric, we plot their signal-to-noise ratios (SNR) [Eq.~\eqref{eq:snr definition}].
    Throughout most of the ergodic regime ($g \gtrsim 0.5$), the SNRs of our metrics and $- \log_{10} D_{\rm KL}$ outperform that of $\langle r \rangle$ by at least an order of magnitude.
    }
    }
    \label{fig:snr cuts}
\end{figure}

Although the level spacing provides an established baseline for distinguishing ergodic and integrable systems, it is not able to resolve finer structure within the ergodic regime.
In particular, Ref.~\cite{Rodriguez-Nieva_2024} showed that by considering the half-chain entanglement entropy of midspectrum eigenstates, one can continuously quantify ergodicity, distinguishing more ergodic systems from less ergodic ones.
In Fig.~\ref{fig:ergodicity maps}(b), we plot this eigenstate entanglement metric $D_{\mathrm{KL}}$ from \cite{Rodriguez-Nieva_2024}, defined as 
\begin{align}
    D_{\mathrm{KL}} &= D^{(1)}_{\mathrm{KL}} + D^{(2)}_{\mathrm{KL}}, \nonumber \\
    D^{(1)}_{\mathrm{KL}} &= \frac{(\mu_E - \mu_R)^2}{2 \sigma_R^2}, \nonumber \\
    D^{(2)}_{\mathrm{KL}} &= \frac{1}{2} \left[ \left( \frac{\sigma_E}{\sigma_R} \right)^2 - 1 \right] - \log \frac{\sigma_E}{\sigma_R} ,
    \label{eq:KL divergence definition}
\end{align}
where $\mu_E$ and $\sigma_E$ are the mean and standard deviation of the half-chain entanglement entropy over the microcanonical ensemble discussed above. 
$\mu_R$ and $\sigma_R$ correspond to the mean and standard deviation of a reference distribution predicted by RMT.
We use the Bianchi-Dona values of $\mu_R = 4.2652$ and $\sigma_R = \sqrt{2}(0.0103)$ for $N = 14$ from Ref.~\cite{Rodriguez-Nieva_2024}.
This metric quantifies the information-theoretic distinguishability of each entanglement distribution and is smaller for more ergodic systems.
% %
Finally, in Fig.~\ref{fig:ergodicity maps}(c) and (d), we plot $\Delta_M$ and $1/D_E$ after averaging over the same microcanonical ensemble.

Comparing each metric in Fig.~\ref{fig:ergodicity maps}, we see that all are able to differentiate the integrable limits on each axis from the ergodic regime in the middle of the plots.
However, only some are able to distinguish more ergodic 
(in the sense that they agree more with the predictions of ETH)
and less ergodic patches within the ergodic regime.
Indeed, the level spacing ratio \edit{saturates to the RMT prediction within this region and fluctuates substantially}, whereas the other three metrics show continuous variations in the same range.
As shown in Ref.~\cite{Rodriguez-Nieva_2024}, the entanglement metric $D_{\mathrm{KL}}$ provides a sensitive continuous ruler that finds a pocket of maximal ergodicity near $g = 1$, $h = 0.3$.
Similarly, both of our metrics $1/D_E$ and $\Delta_M$ peak in the same region of parameter space and are therefore able to identify maximal ergodicity.
Furthermore, our metrics are well-defined and non-singular across all of parameter space, in contrast with $D_{\mathrm{KL}}$ and $\langle r \rangle$, which diverge and show \edit{larger} fluctuations respectively for $g=0$.
These results confirm that the robustness of Hamiltonian learning not only indicates ergodicity but also \textit{quantifies} it.

\edit{
To gain a more quantitative understanding of how sensitive each metric is to a Hamiltonian's level of ergodicity, we plot line cuts of the data in Fig.~\ref{fig:ergodicity maps} at $h = 0.3$ in Fig.~\ref{fig:snr cuts}(a-b).
Each metric is also plotted with a ribbon indicating the standard error of each metric.
Because $\Delta_M$ and $1/D_E$ ($\langle r \rangle$) are averaged over different eigenstates (sets of three adjacent eigenstates), their standard errors are computed from the standard deviation over the ensemble.
For $- \log_{10} D_{\rm KL}$, we bootstrap the standard error by resampling with replacement $5000$ times.
It's clear that while the level spacing suffers from large eigenstate-to-eigenstate fluctuations, the other three metrics vary smoothly with minimal noise.
}

\edit{
Furthermore, we can quantify the sensitivity of each metric $f(g)$ by computing its signal-to-noise ratio (SNR) as a function of the parameter $g$:
\begin{align}
    \mathrm{SNR}_f(g) = \frac{|\partial_g f(g)|}{\sigma_f(g)},
    \label{eq:snr definition}
\end{align}
where $\sigma_f(g)$ is the standard error in $f$ at $g$ and $|\partial_g f(g)|$ quantifies the ability of $f$ to detect changes in ergodicity as one varies $g$.
However, this definition raises a subtlety: how do we separate the noise from the signal in the data in Fig.~\ref{fig:snr cuts}?
For $\Delta_M$, $1/D_E$, and $-\log_{10} D_{\rm KL}$, where the noise is generally an order of magnitude smaller than the point-to-point change, we can safely approximate the signal by the plotted means.
However, for the level spacing, the point-to-point changes are evidently dominated by the noise.
As a proxy for the ``true'' signal, we use the level spacing $\langle r \rangle$ calculated for the \textit{entire} energy spectrum rather than 600 eigenstates since the former has a far lower standard error (for more details, see Appendix~\ref{app:metric sensitivity details}).
Computing the numerator with this value and the denominator with the noise in Fig.~\ref{fig:snr cuts}(a), we get the SNR values shown in Fig.~\ref{fig:snr cuts}(c).
Clearly, the SNR of $\Delta_M$, $1/D_E$, and $-\log_{10} D_{\rm KL}$ remains at least an order of magnitude above that of $\langle r \rangle$ throughout much of the ergodic regime ($g \gtrsim 0.5$), providing quantitative evidence that these metrics can more sensitively probe ergodicity due to their smaller relative fluctuations.
For more details on the level spacing and the system size dependence of these metrics, see Appendix \ref{app:metric sensitivity details}.
}

\section{Probing Eigenstate Sensitivity}
\label{sec:probing chaos}

\begin{figure}[t]
    \centering
    \includegraphics[width=\linewidth]{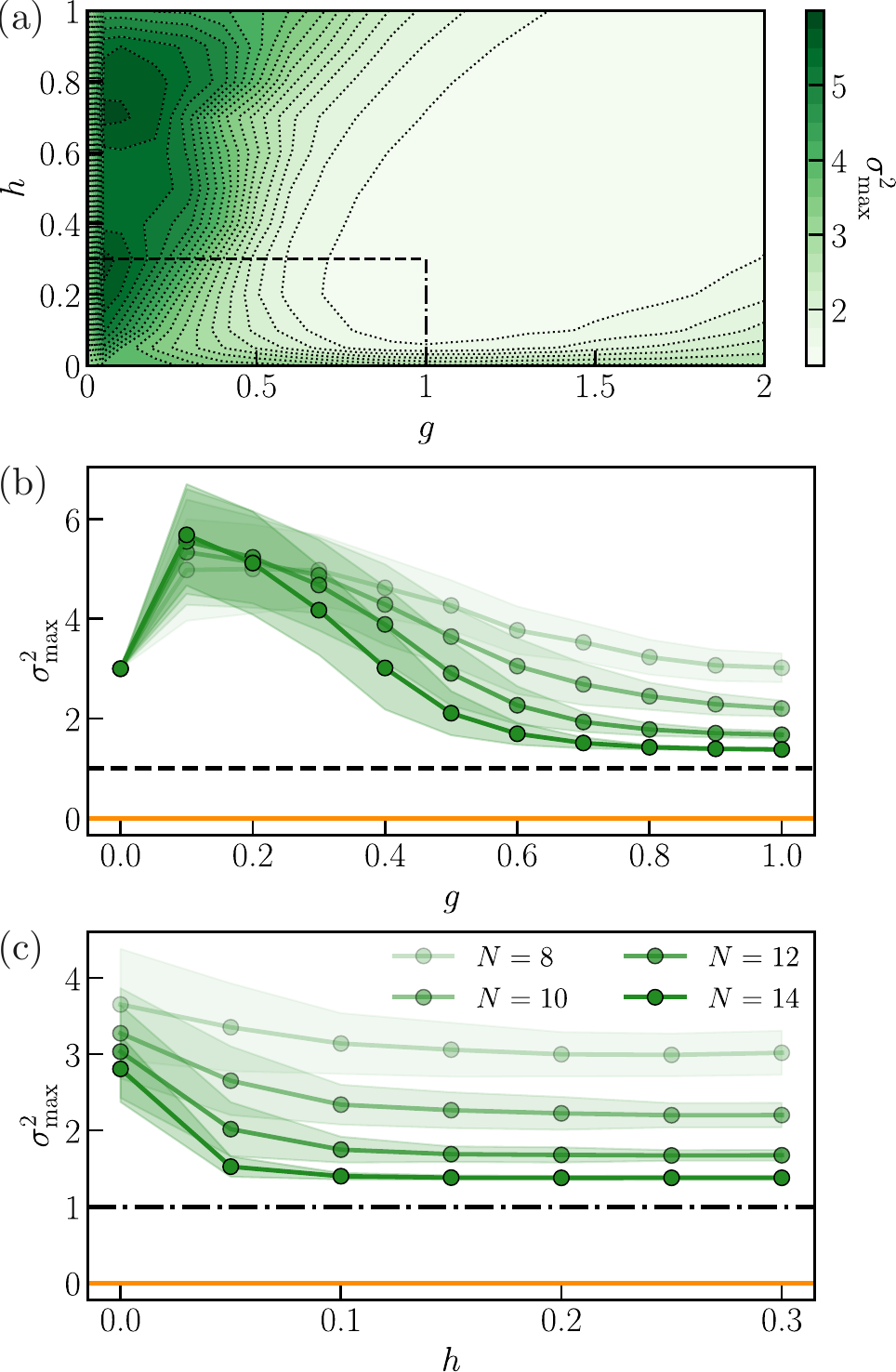}
    \caption{
    \textbf{Probing Chaos via Eigenstate Sensitivity.}
    We compute the average maximal variance of the variance spectra from Fig.~\ref{fig:ergodicity maps} as a probe of the sensitivity of eigenstates to local perturbations.
    (a) This metric identifies a window of parameter space near $g \to 0$ for $N = 14$ where eigenstates have high sensitivity, separating integrable and ergodic limits \cite{Pandey_2020,LeBlond_2021,Kim_2025} despite looking nearly integrable in Fig.~\ref{fig:ergodicity maps}.
    (b) If we furthermore consider a slice of this parameter space ($h=0.3$), we see a transient increase of $\sigma_{\mathrm{max}}^2$ with system size in the sensitive region before it shrinks as the ergodic region expands.
    (c) In contrast, we observe no such scaling between the ergodic region and the transverse field limit ($g=1$).
    We attribute this lack of sensitivity to the strict locality of the operators $\mathcal{O}_a$.
    Ribbons denote the standard deviation of $\sigma^2_{\mathrm{max}}$ over the ensemble of states.
    }
    \label{fig:chaos map}
\end{figure}

While the study of quantum chaos often focuses on ergodicity and the predictions of ETH, it also considers phenomena like the sensitivity of eigenstates to perturbations (encoded in the QFI and related measures).
As such, we now turn to understanding the variance spectrum's ability to probe chaos more broadly, revealing structure in parameter space unseen by the metrics in Fig.~\ref{fig:ergodicity maps}.

In particular, previous work defines ``maximally chaotic'' regions of parameter space as those in which the AGP norm saturates its maximum possible growth with system size \cite{Pandey_2020,LeBlond_2021,Kim_2025}, indicating that the average eigenstate is maximally sensitive to a small change in its parent Hamiltonian.
The AGPs used to detect this behavior are very nonlocal compared to the Hamiltonian \cite{Pandey_2020}, and as such, the variance spectrum is not able to detect the onset of true maximal chaos as in these previous works.
However, it is still able to quantify the sensitivity of the given eigenstate to all \edit{geometrically} $\ell$-local AGPs and can identify regions of parameter space where this restricted type of sensitivity is maximized.

We will now see that this ability reveals structure in parameter space not seen by the metrics in Fig.~\ref{fig:ergodicity maps}, uncovering a section of parameter space that behaves qualitatively differently from both the integrable and ergodic limits.
To probe such behavior, we consider the maximal variance $\sigma_{\mathrm{max}}^2$ (or $\sigma_{N_L}^2$) of the variance spectrum.
As stated in Section~\ref{sec:analytic results}, $1/\sigma_{\mathrm{max}}^2$ \edit{is proportional to} the minimum achievable Hamiltonian susceptibility $\chi_H$ \eqref{eq:chi H}.
We will now show how this metric identifies regions of parameter space where this eigenstate sensitivity to local perturbations is not only maximized but also grows with system size.
This behavior is reminiscent of maximal chaos and, more importantly, uncovers structure in the energy spectrum unseen by the metrics in Fig.~\ref{fig:ergodicity maps} that identify this region as nearly integrable.

In Fig.~\ref{fig:chaos map}(a), we plot this maximal eigenvalue (averaged over the microcanonical ensemble) in the MFIM for the same parameter choices as Fig.~\ref{fig:ergodicity maps}.
While the ergodic regime has $\sigma_{\mathrm{max}}^2 \approx 1$ as expected, this value does not peak in the integrable limits.
Rather, we see an intermediate region ($g \to 0$, nonzero $h$) separating the ergodic and classical regimes where this probe of chaos peaks.
Furthermore, Fig.~\ref{fig:chaos map}(b)---where we plot a $h = 0.3$ slice of (a) for various $N$ values---shows that $\sigma_{\mathrm{max}}^2$ not only peaks in this region but has a transient \textit{increase} in system size.
This is in stark contrast to all other regions of parameter space plotted: in the classical limit ($g=0$), $\sigma_{\rm max}^2$ is constant, whereas both the free fermion line ($h=0$) and the ergodic regime have strictly decreasing values of $\sigma_{\mathrm{max}}^2$.
Although this increase reverses once $N$ becomes large enough and the ergodic region expands in parameter space, this behavior is consistent with previous works which find that the maximally chaotic region shrinks as $N$ increases \cite{Pandey_2020,LeBlond_2021,Kim_2025}.
Note that this enhanced eigenstate sensitivity is invisible to most metrics of ergodicity, as this region looks nearly integrable in Fig.~\ref{fig:ergodicity maps}.

This high QFI can be understood via analogy with GHZ states \cite{greenberger1989going}. 
Indeed, the eigenstates for $g=0$ correspond to bitstrings in the $Z$ basis which will be degenerate if they have the same magnetization and number of domain walls.
Upon breaking integrability by increasing $g$, the transverse field will generally hybridize degenerate states and lift the degeneracy.
These GHZ-like superpositions of bitstring states have a high QFI for $Z$ basis observables, and our numerics confirm that the operators $\mathcal{O}_{\mathrm{max}}$ with the highest variance $\sigma_{\mathrm{max}}^2$ are primarily composed of $Z$ operators in the \edit{maximally} sensitive region.

In contrast, Fig.~\ref{fig:chaos map}(c) shows no such region of enhanced sensitivity separating the free fermion integrable limit ($h=0$) and the ergodic regime for the slice $g = 1$.
However, maximal chaos \textit{has} been observed numerically in this region when studying the norm of the AGP \cite{Pandey_2020}.
We attribute this discrepancy to the fact that we only probe the QFI of \edit{geometrically} $\ell$-local generators whereas AGPs are generally nonlocal.
Indeed, such distinctions between local and nonlocal AGPs have also been observed when considering adiabatic flows in the MFIM \cite{Sugiura_2021,Kim_2024}.
This difference points to the drawbacks of considering \edit{geometrically} $\ell$-local AGPs when probing eigenstate sensitivity as we note in Table~\ref{tab:metric comparison}.

\section{Experimental Implementation and Robustness}
\label{sec:state prep and measurement}

\begin{figure}[t]
    \centering
    \includegraphics[width=\linewidth]{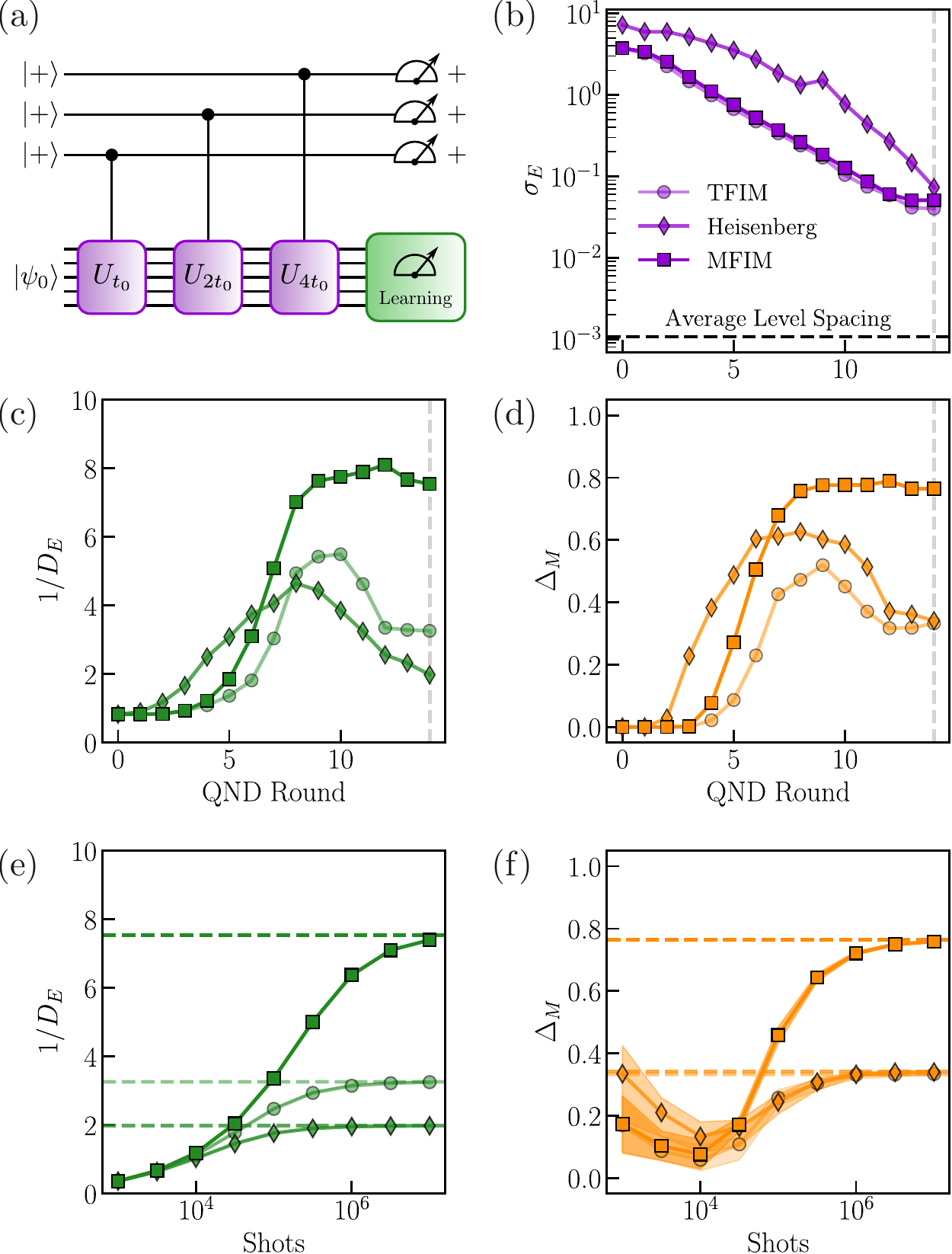}
    \caption{
    \textbf{Approximate Eigenstate Preparation and Measurement.}
    (a) We sketch the quantum non-demolition (QND) algorithm for preparing an approximate eigenstate from Ref.~\cite{Vasilyev_2020}. 
    (b) Starting from the bitstring state $\ket{\psi_0}$ with $\langle H \rangle$ closest to 0, the controlled evolution and post-selection on auxiliary qubits reduce the state's energy variance $\sigma_E^2$ exponentially in the number of QND measurements.
    However, the prepared states still have energy width much larger than the average level spacing plotted in black (see text for details).
    The integrable models show a peak in (c) $1/D_E$ and (d) $\Delta_M$ at intermediate energy variances before approaching their eigenstate values from Fig.~\ref{fig:vars vs energy}. 
    Using the states prepared after 14
    %8
    QND rounds (gray dashed line), we perform random local Pauli measurements \cite{Huang_2020} to efficiently extract all correlation functions necessary to calculate (e) $1/D_E$ and (f) $\Delta_M$. Although the latter shows significant fluctuations for few shots, the former monotonically increases with minimal fluctuations. Ribbons denote the standard deviation over 10 repetitions and colored dashed lines indicate the exact values of each metric. 
    }
    \label{fig:eigenstate preparation}
\end{figure}

Although we have shown that our metrics are able to probe ergodicity and chaos in theory and numerical simulations, we also desire practical metrics that can be implemented in experiments on quantum simulators.
Due to the locality of the observables $L_{\alpha}$, the variance spectrum can be efficiently measured in such settings as long as midspectrum energy eigenstates can be prepared.
Unfortunately, this requirement is generally difficult to achieve, as the preparation requires time scaling exponentially in system size even with access to universal quantum computers~
\cite{Atia_2017,Abrams_1999,Poulin_2009,Lin_2020}.
However, many proposed methods allow for the preparation of \textit{approximate} midspectrum eigenstates that have a small but nonzero energy variance \cite{Yang_2020,Vasilyev_2020,Irmejs_2024,Garratt_2024}.
\edit{
Indeed, given our previous discussion of ETH, one would expect that reducing a state's energy width to below the Thouless energy, not the Heisenberg energy (after which one has effectively prepared an eigenstate), is sufficient to detect the onset of ergodicity.
}
To this end, we now demonstrate how our metrics behave under such preparation schemes and how the covariance matrix can be efficiently extracted using randomized measurements \cite{Huang_2020}.

\edit{
It is worth noting that even if one does not know the Hamiltonian of the system in question, common access models used in Hamiltonian learning allow for access to e.g. black box time evolution under $H$ \cite{Lindner-Bairey_2019,Bentsen_2019,Hsieh-Li_2020,Hu_2025}.
As we will see, this is generally sufficient to prepare approximate midspectrum eigenstates.
As such, one is able to both learn a system's Hamiltonian and also quantify how ergodic it is in one protocol.
However, we also emphasize that even if one already knows the Hamiltonian of a system, our metrics are still useful as general probes of chaos and ergodicity.
}

\subsection{State Preparation}

In Fig.~\ref{fig:eigenstate preparation}(a), we show a diagram of the quantum non-demolition (QND) algorithm from Ref.~\cite{Vasilyev_2020} for preparing approximate eigenstates.
Although other algorithms \cite{Lin_2020,Garratt_2024} exist which offer advantages over this phase estimation-based method, we use this simple approach due to its generality and direct relevance to existing experimental platforms such as Rydberg atom arrays \cite{Vasilyev_2020}.
Starting from our system of interest (prepared in some initial state $\ket{\psi_0}$) along with an auxiliary register of $r$ qubits (prepared in the state $\ket{+}^{\otimes r}$), the algorithm performs unitary time evolution under our Hamiltonian of interest $H$ controlled by each of our auxiliary qubits.
Specifically, we first evolve under $U_{t_0} = \exp[-i H t_0]$ for time $t_0 = \pi / ||H||_{\infty}$ controlled by the first auxiliary qubit.
We then evolve for time $2 t_0$ controlled by the second qubit, $4 t_0$ by the third qubit, and so on, doubling the evolution time for each successive qubit.
We then projectively measure all auxiliary qubits in the $\ket{\pm}$ basis with each successive measurement corresponding to a single QND round.

By post-selecting to ensure that all auxiliary qubits end in the state $\ket{+}$, the system's final state will have energy near $0$ with an exponentially suppressed energy variance compared to $\ket{\psi_0}$.
While the success probability of this procedure does decay exponentially with the number of QND rounds, we will show that the number of rounds required to adequately measure ergodicity is substantially lower than the number required to prepare a true eigenstate.
As such, the state preparation protocol for our metrics has a success probability exponentially larger than that of preparing an eigenstate.
\edit{
Indeed, we do not expect to observe any clear indicators of ergodicity before the state's energy width falls below the Thouless energy, so this sets the minimum number of QND rounds required for any metric of ergodicity.
}
Furthermore, note that even if we do not post-select and instead consider \textit{all} runs regardless of the measurement outcome, each state in the resulting ensemble will still have low energy variance but need not have average energy $\langle H \rangle \approx 0$.
Rather, the energies of the states will be distributed according to the width of the initial state $\ket{\psi_0}$ in energy.
Because we will choose $\ket{\psi_0}$ to be a product state with $\bra{\psi_0} H \ket{\psi_0} \approx 0$, it furnishes a thermal distribution with an effective temperature $T \to \infty$ and will in general converge to the predictions of an infinite temperature microcanonical ensemble as $N \to \infty$.
As such, averaging our ergodicity metrics over all measurement outcomes is still able to probe ergodicity without post-selection.

In Fig.~\ref{fig:eigenstate preparation}(b), we consider the TFIM, Heisenberg model, and MFIM with $N = 14$ qubits and initialize each system in the $Z$ bitstring state with average energy $\bra{\psi_0} H \ket{\psi_0}$ closest to 0.
We then perform successive controlled evolutions and projective QND measurements into the auxiliary $\ket{+}$ state.
As expected, the energy variance with respect to the exact Hamiltonian $\sigma_E^2$ decreases exponentially with the number of QND rounds.
Notice, however, that even the state prepared after 14 QND rounds has an energy width $\sigma_E$ nearly 2 orders of magnitude larger than the average level spacing plotted in black.
This value is calculated as the level spacing $\Delta E_n$ averaged over 600 eigenstates centered at $E=0$ in each model, and although we only plot the largest of the three values, all three models have nearly the same average level spacing.\footnote{All three models have average level spacing $8 \times 10^{-4} < \langle \Delta E_n \rangle < 1.1 \times 10^{-3}$.}
After each round, we also numerically calculate the inverse spread $1/D_E$ [Fig.~\ref{fig:eigenstate preparation}(c)] and gap $\Delta_M$ [Fig.~\ref{fig:eigenstate preparation}(d)] of the covariance matrix of the prepared state.
\edit{
Note that, although $M$ will in general have no zero eigenvalues for an approximate eigenstate (except for the Heisenberg model, where the magnetization will always have zero variance), we can still define the gap $\Delta_M$ as in Eq.~\eqref{eq:Delta_M definition}, where $\sigma_0^2$ can be nonzero.
}
% Also note that at these intermediate stages, $M$ may have no zero eigenvalues since the state need not be an eigenstate of any local operator
% (except for the magnetization in the case of the Heisenberg model).
%
Although all three systems start with identical variance spectra, the metrics' behavior depends on the integrability of each system.
While both metrics increase almost monotonically in the MFIM before saturating near their values from Fig.~\ref{fig:vars vs energy}, they show a peak at intermediate stages of the preparation in the TFIM and Heisenberg models before decreasing toward their eigenstate values.
The projection of the product state into a small but non-negligible window of the energy spectrum is therefore able to increase both $\Delta_M$ and (to a lesser extent) $1/D_E$ beyond their values for eigenstates, but this increase does not overcome the separation between ergodic and integrable regimes.
%
% \edit{
% \sout{Also note that at these intermediate stages, $M$ may have no zero eigenvalues since the state need not be an eigenstate of any local operator (except for the magnetization in the case of the Heisenberg model).}
% }
%
Furthermore, our metrics are still able to easily and quantitatively distinguish integrable and ergodic regimes for energy fluctuations $\sigma_E \sim 10^{-1}$ (after $>10$ QND rounds), orders of magnitude larger than the midspectrum level spacing.
As such, approximate eigenstates
are sufficient to distinguish ergodic and integrable systems using the variance spectrum.

\subsection{Measurement}

Having considered approximate eigenstate preparation, we now turn to efficiently extracting the covariance matrix from the states prepared by this algorithm.
We will begin by discussing the sample complexity of measuring the covariance matrix $M$ such that any individual element is known to additive error $\epsilon$.
We will then discuss what $\epsilon$ is needed to resolve the different aspects of ergodicity and chaos discussed above.

Because all expectation values that appear in $M$ have weight $\leq 2\ell$, they can be efficiently extracted in parallel using random local Pauli bases according to classical shadow tomography \cite{Huang_2020,Zhan_2024,Aaronson_2018,Elben_2022}.
In particular, we take the state prepared after 14
% 8 
QND measurements (gray dashed line) and measure each qubit in the $X$, $Y$, or $Z$ basis at random. 
This measurement will result in a product state where each qubit is in one of the states $\{ \ket{0}, \ket{1}, \ket{+}, \ket{-}, \ket{+i}, \ket{-i}\}$.
If we repeat this procedure $N_s$ times, we obtain $N_s$ classical bitstrings encoding the measurement outcomes of $N_s$ random choices of bases.
These measurements can then be used to classically construct an estimator of the density matrix of the original state:
\begin{align}
    \hat{\rho} = \frac{1}{N_s} \sum_{n_s = 1}^{N_s} &(3 |s_1^{(n_s)}\rangle \langle s_1^{(n_s)}| - I) \otimes \nonumber \\
    &\cdots \otimes (3 |s_N^{(n_s)}\rangle \langle s_N^{(n_s)}| - I), 
\end{align}
where $|s_i^{(n_s)}\rangle$ is the state of qubit $i$ after the $n_s$th measurement.
This estimator is sufficient to learn the expectation values of $N_M$ $2\ell$-local operators to additive error $\epsilon$ using just $N_s = O(3^{2\ell} \log(N_M)/\epsilon^2)$ shots.
\edit{
Note that these operators need not be geometrically local; their weight is the sole attribute determining their measurement complexity.
}
Notably, $N_s$ only scales logarithmically in system size, since $N_M = O(N_L^2) = O(N^2)$.
We note that this efficiency 
follows from the locality of our basis $L_{\alpha}$.
In contrast, measuring nonlocal objects like the half-chain entanglement entropy 
with full state tomography generally requires a number of shots scaling exponentially with $N$.

We demonstrate the shadows measurement scheme in Fig.~\ref{fig:eigenstate preparation}(e) and (f) for the state prepared after 14 QND rounds, where we plot the extracted values of $1/D_E$ and $\Delta_M$ respectively for each model as a function of the number of shots $N_s$.
For $N=14$, $N_s \sim 10^6$ suffices to learn all $N_M = 12720$ unique matrix elements of $M$ to reasonably converge to their exact values (represented as colored dashed lines).
Although this value of $N_s$ may seem large given the $\log N$ scaling discussed above, it is still consistent with the prefactor imposed by the scaling of shadow tomography.\footnote{Although derandomized shadows are often used to reduce this constant prefactor, our preliminary investigations using Ref.~\cite{Huang_2021} showed essentially no improvement. We attribute this to the large size of our operator basis conditioned only by geometric locality.}
For instance, if we demand that all matrix elements are learned to an error $\epsilon = 0.01$, one can estimate the requisite $N_s$ as $3^{2\ell} \log(N_M)/\epsilon^2 \approx 7 \times 10^6$. 
Note that while $\Delta_M$ shows large fluctuations for low values of $N_s$ and changes non-monotonically as we increase the number of shots, $1/D_E$ shows almost no fluctuations for any $N_s$ and increases monotonically with the number of shots.
This suggests that measuring large values of $1/D_E$ can accurately witness ergodicity even with limited $N_s$, but we leave a full analysis of this behavior to future work.

Given this scaling, one must then ask how $\epsilon$ scales with $N$ to reach, say, a desired additive error in $D_E$.
For instance, in the ergodic regime, the exponential decay of $D_E$ with $N$ suggests that locating regions of maximal ergodicity requires knowing this metric to an exponentially small additive error, requiring an exponential number of shots.
However, if our goal is instead to continuously quantify ergodicity and chaos to inverse polynomial error---such that we can, e.g., observe quantitative differences between integrable and ergodic systems and detect regions of enhanced eigenstate sensitivity---the required $\epsilon$ will only scale polynomially in system size (see Appendix~\ref{app:noise error bounds} for a more explicit argument for this scaling).
As such, the variance spectrum still provides a route to quantifying ergodicity and chaos with a polynomial number of measurements $N_s$.

\section{Discussion and Outlook}

In this work, we have established new connections between the study of quantum chaos and ergodicity and the theory of Hamiltonian learning, using the latter to accurately quantify the former.
In particular, the variance spectrum \eqref{eq:var spectrum} can both distinguish integrable and ergodic regimes and quantify ergodicity within the latter regime. 
Beyond ergodicity, the variance spectrum is also able to probe chaos via the sensitivity of eigenstates to local perturbations.
These abilities allow us to uncover detailed maps of parameter space [Fig.~\ref{fig:ergodicity maps}(c-d) and Fig.~\ref{fig:chaos map}(a)] in the MFIM and identify regions of maximal ergodicity and enhanced local sensitivity.
Furthermore, the locality of these observables allows us to efficiently measure the variance spectrum from experimentally-preparable approximate energy eigenstates using the classical shadows formalism.

Our work motivates a number of future directions.
First, metrics like $D_E$ may benefit from more precise predictions of their behavior in the ergodic regime.
Utilizing reference distributions at finite system sizes \cite{Sierant_2019} rather than predictions for the thermodynamic limit [as in Eq.~\eqref{eq:D_E definition}] may increase the sensitivity of this metric in the ergodic regime, similar to the sensitivity of $D_{\mathrm{KL}}$ in Fig.~\ref{fig:ergodicity maps}(b).
Furthermore, Ref.~\cite{Rodriguez-Nieva_2024} uses the Bianchi-Dona distribution as a reference for eigenstate entanglement rather than the Page distribution because it accounts for the constraint of energy conservation and correctly predicts $O(1)$ deviations from Haar random distributions.
Indeed, understanding to what extent midspectrum eigenstates can be modeled as Haar random has implications across all studies of quantum ergodicity.
Our metrics could therefore benefit from a more thorough analysis of the effects of energy conservation.
Although we perform preliminary investigations in this direction in Appendix~\ref{app:krylov overlaps}, we leave a detailed analysis to future work.

Second, integrable systems without disorder are not the only models that avoid thermalization.
While the effect of localization in disordered systems on the gap $\Delta_M$ has been investigated before using a different operator basis \cite{Dupont_2019,Dupont_2019_2}, our results also point to the utility of analyzing $1/D_E$ and $\sigma_{\mathrm{max}}^2$ as a way to probe the transition between ergodic, chaotic, and localized regimes \cite{Sels_2021}.
Furthermore, the effect of other ergodicity-breaking mechanisms on the variance spectrum remains to be explored, including quantum many-body scars \cite{Bernien_2017,Turner_2018,Turner_2018_2}, Hilbert space fragmentation \cite{Sala_2020,Khemani_2020}, and more exotic forms of integrability \cite{Fendley_2019,Gombor_2021}.
Moreover, the AGP norm has been used as a metric for chaos in classical systems as well \cite{Kim_2025,Karve_2025}.
Understanding whether the variance spectrum can be applied to e.g. stationary ensembles in classical systems could generalize this metric beyond the quantum regime.

Third, our work motivates a deeper analytic understanding of small gaps in the variance spectrum.
In particular, understanding how much of our free fermion derivation in Appendix~\ref{app:ff predictions} generalizes to Bethe integrable systems remains an interesting open question.
Indeed, the fact that our method relies on the quasiparticle nature of the spectrum would seem to indicate that it can generalize to many other forms of ergodicity breaking \cite{Chandran_2023,moudgalya2023exhaustivecharacterizationquantummanybody,Korepin_Bogoliubov_Izergin_1993}.
Furthermore, our analytic approach shares many similarities with ground state physics in gapless systems, where conformal field theory and Lieb-Schultz-Mattis theorems take center stage.
Although conformal field theory can explain results for half-filling and similar ``ground states'' in the middle of the free fermion spectrum, we have also shown its utility in bounding $\Delta_M$ for general states in the spectrum.
Understanding how far this analogy between gapless phases of matter and integrability extends represents an important line of work and includes studies of e.g. renormalization group-like flows from AGP norms \cite{Sugiura_2021,Kim_2024}.

In the present work, we have only focused on Hamiltonian learning methods that use an eigenstate as a resource, yet our work motivates the investigation of whether other learning algorithms can quantify ergodicity and chaos.
In particular, many Hamiltonian learning algorithms use quench dynamics or finite-temperature Gibbs states as a resource \cite{Lindner-Bairey_2019,Hsieh-Li_2020,Bentsen_2019,Anshu_2021,bakshi2023learningquantumhamiltonianstemperature}. 
Understanding how these methods depend on ergodicity could lead to metrics that do not require eigenstate preparation.
Furthermore, ETH predicts that states obtained after long-time quench dynamics under an ergodic Hamiltonian still look thermal when restricted to local subsystems.
As such, the structure of the variance spectrum after such evolution should still resemble that of an eigenstate with the exception that the Hamiltonian will have nonzero variance.
It may therefore be possible to still use the variance spectrum as a metric for ergodicity and chaos despite not having access to an (approximate) eigenstate, but we leave these questions to future work.

\section*{Acknowledgements}

We thank Samuel Buckley-Bonanno,
Sarang Gopalakrishnan,
Vedika Khemani, Daniel Mark, Stewart Morawetz, Anatoli Polkovnikov, Nikita Romanov, and Rahul Sahay for illuminating discussions, and we thank Hong-Ye Hu and Anatoli Polkovnikov for a careful reading of this manuscript.
We especially thank Anatoli Polkovnikov for pointing out ways that the AGP norm can be generalized to map onto the variance spectrum.
We also thank Harvard University's FAS Research Computing for numerical resources.
We thank the NSF for funding through the CUA PFC (PHY-2317134) and QuSEC (OMA-2326787) as well as the AFOSR (FA9550-24-1-0311). 
C.K. acknowledges funding from the NSF through a grant at ITAMP.
S.C. also acknowledges support from the NSF CAREER Award DMR-2237244, the Heising-Simons Foundation (grant \#2024-4851), and the Alfred P. Sloan Fellowship.

%\bibliography{refs.bib}


\appendix

\edit{
\section{Choice of Boundary Conditions}
\label{app:boundary conditions}
}

\begin{figure*}[t]
    \centering
    \includegraphics[width=\linewidth]{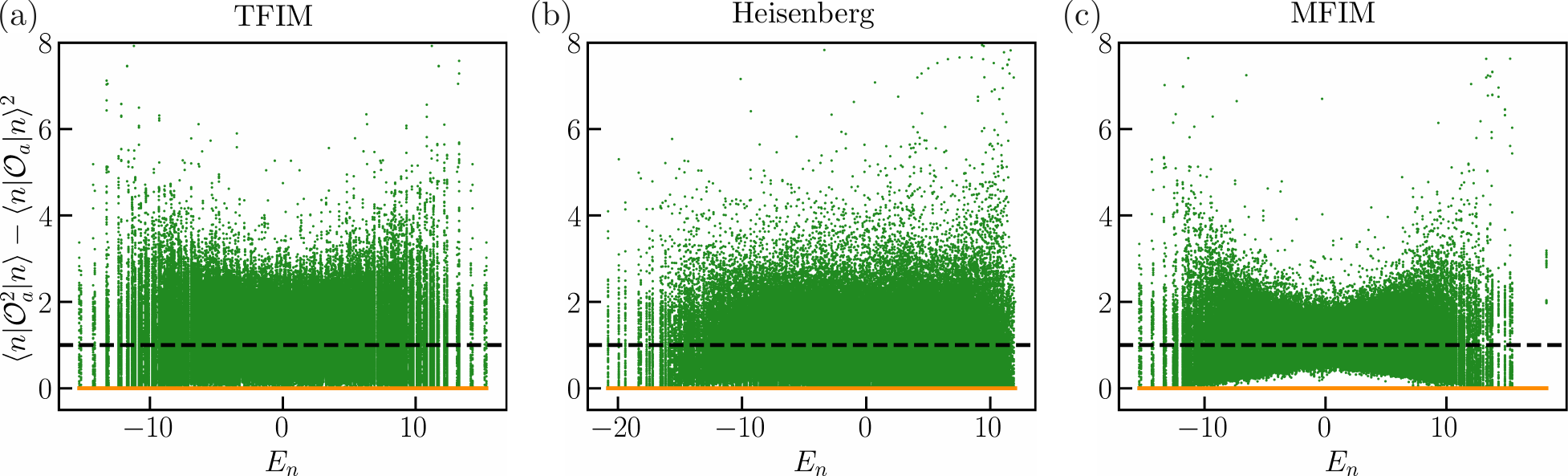}
    \caption{
    \edit{
    \textbf{Variance Spectra with Different Boundary Conditions.}
    We plot the variance spectra for the same models as in Fig.~\ref{fig:vars vs energy}(a-c) with the simple change that all boundary fields are set to zero and the chain is made periodic.
    As expected, the general behavior of the variance spectrum across different energies is largely the same as in Fig.~\ref{fig:vars vs energy}.
    Although the variance spectrum of the MFIM is less tightly confined near $E=0$, we attribute this to the smaller available Hilbert space any given eigenstate can hybridize with due to translation and inversion symmetry.
    }
    }
    \label{fig:boundary conditions}
\end{figure*}

\edit{
In the main text, we choose to consider 1D chains of qubits with open boundary conditions and various small fields applied to the edges of the chains.
This choice not only allows us to compute metrics like $\langle r \rangle$ without first restricting to a single symmetry sector but also effectively grows the available Hilbert space over which eigenstates can hybridize in the ergodic limit, reducing RMT fluctuations.
In this appendix, we verify that this choice is not essential to our arguments and discuss details of imposing translation invariance on the Hamiltonian ansatz.
}

\edit{
We first numerically verify in Fig.~\ref{fig:boundary conditions} that the variance spectra plotted in Fig.~\ref{fig:vars vs energy} do not change dramatically if we (1) remove the additional boundary fields (setting $h_1 = h_N = 0$) and (2) make the chains periodic. 
Even with this modification, Fig.~\ref{fig:boundary conditions} shows the same general features as those in the main text: the integrable models' variance spectra [in panels (a-b)] have small gaps and large spreads across all values of $E$ whereas the non-integrable MFIM has a concentrated variance spectrum [in panel (c)] with a large gap as one approaches $E=0$.
We note that the variance spectrum in the periodic MFIM is not as tightly concentrated as the open MFIM in the text.
We attribute this to the larger fluctuations due to the smaller available Hilbert space in the periodic case.
}

\edit{
One might also wonder how imposing symmetries like translation invariance on the Hamiltonian ansatz (e.g. the operator basis $\{L_{\alpha}\}$) affects our results.
Here, the answer is more subtle; as we discuss in Section~\ref{sec:analytic results} and Appendix~\ref{app:ff predictions}, the small gaps in translation-invariant free fermion systems arise from operators that are not translation invariant.
Rather, they are constructed by taking the translation-invariant Hamiltonian and multiplying it by a long-wavelength sinusoid in space.
As such, restricting the operator basis to only include operators like $\sum_i L_{i \mu}$---where $i$ denotes the location of the operator and $\mu$ denotes the species of operator (e.g. $X$, $Y$, or $Z$)---will explicitly exclude these low variance operators and artificially inflate the gap in the variance spectrum.
}

\edit{
However, this argument does not imply that symmetry cannot be enforced in any way on the operator basis.
Indeed, Ref.~\cite{Qi_2019} points out that translation invariance allows one to block diagonalize the covariance matrix $M$ into different momentum sectors labeled by $k$.
In their treatment, one rotates the basis $L_{i \mu}$ using a Fourier transform given by
\begin{align}
    \widetilde{L}_{k\mu} = \frac{1}{\sqrt{N}} \sum_j L_{j \mu} e^{i k j}
\end{align}
where $k$ denotes the momentum of the operator (e.g. $k=0$ means $\widetilde{L}_{k\mu}$ is translation invariant).
Ref.~\cite{Qi_2019} then explicitly shows that, for a translation-invariant state $\ket{v}$, $M$ is block diagonal in this basis such that
\begin{align}
    \widetilde{M}_{k\mu,p\nu} = \delta_{kp} \widetilde{f}_{\mu \nu} (k),
\end{align}
 where $\widetilde{f}$ is some Fourier-transformed correlation function, and comments that this can be extended to states $\ket{v}$ that have some nonzero momentum themselves.
Thus, while enforcing translation invariance on the learned Hamiltonian corresponds to restricting to the $k=0$ block of $\widetilde{M}$, simply block diagonalizing the matrix allows one to enforce translation symmetry on $M$ without artificially inflating its gap.
}

\section{RMT Calculations for Variance Spectra}
\label{app:rmt predictions}

To estimate the variance spectrum of infinite-temperature eigenstates in the ergodic regime, we assume such states act as Haar random with respect to all local observables orthogonal to $H$ (such operators will still generally have nonzero overlap with the conserved operators $H^2, H^3, \dots$, but we delay this discussion to Appendix~\ref{app:krylov overlaps}).
This allows us to use the machinery of random unitaries \cite{Mele_2024} to make predictions in the ETH regime.
Although real Hamiltonians (with time reversal symmetry) will have real eigenvectors and therefore will not generate a Haar random ensemble of eigenstates, we focus on the more general Haar random case for simplicity. 
Moreover, this caveat generally only incurs a constant scale factor correction in the resulting predictions for e.g. entanglement entropy \cite{Rodriguez-Nieva_2024,Vivo_2016,Wei_2017,Lukin_2019}.

If we define $\mathcal{U}$ as the ensemble of Haar random unitaries, we can exchange our eigenstate $\ket{v}$ for a Haar random state $U\ket{0}$ where $U \sim \mathcal{U}$ and $\ket{0}$ is an arbitrary 
reference state.
Then the expected value of a local operator $\mathcal{O}_a$'s variance is given by
\begin{align}
    \mathbb{E}[\sigma_a^2] = \mathbb{E}[\bra{0} U^{\dagger} \mathcal{O}_a^2 U \ket{0} ] - \mathbb{E}[\bra{0} U^{\dagger} \mathcal{O}_a U \ket{0}^2 ] .
\end{align}
Using the first moment of the Haar measure \cite{Mele_2024}, the first term is given by
\begin{align}
    \mathbb{E}[\bra{0} U^{\dagger} \mathcal{O}_a^2 U \ket{0} ] = \frac{1}{d} \Tr[\mathcal{O}_a^2] = 1,
\end{align}
where we have used the fact that all $\mathcal{O}_a$ under consideration are normalized.
The second term uses the second moment of the Haar measure:
\begin{align}
    \mathbb{E}[\bra{0} U^{\dagger} \mathcal{O}_a U \ket{0}^2 ] &= \frac{1}{d(d+1)} (\Tr[\mathcal{O}_a]^2 + \Tr[\mathcal{O}_a^2]) \nonumber \\
    &= \frac{1}{d+1} ,
\end{align}
giving a final expectation value of
\begin{align}
    \mathbb{E}[\sigma_a^2] = 1 - \frac{1}{d+1} .
    \label{eq:rmt mean of var}
\end{align}
Although this calculation depends on the Hilbert space dimension $d$, the contribution that introduces this dependence decreases exponentially in the system size $N$.
Indeed, even for the modestly sized systems considered in this work ($N = 14$), this contribution is already less than $10^{-4}$.
As such, the mean of the variance spectrum can be well approximated by $1$ in the ergodic regime.

Let us now consider the fluctuations around 1, which are given by $\mathbb{E} [(\sigma_a^2 - 1)^2] = \mathbb{E} [(\sigma_a^2)^2] - 2 \mathbb{E} [\sigma_a^2] + 1$.
The second term is provided to us by Eq.~\eqref{eq:rmt mean of var}, while the first term in this expression becomes
\begin{align}
    \mathbb{E} [(\sigma_a^2)^2] &= \mathbb{E}[\bra{0} U^{\dagger} \mathcal{O}_a^2 U \ket{0}^2 ] \nonumber \\
    &- 2 \mathbb{E}[\bra{0} U^{\dagger} \mathcal{O}_a^2 U \ket{0} \bra{0} U^{\dagger} \mathcal{O}_a U \ket{0}^2 ] \nonumber \\
    &+ \mathbb{E}[ \bra{0} U^{\dagger} \mathcal{O}_a U \ket{0}^4 ] .
\end{align}
The first term in this sum can be calculated using the second moment of the Haar measure:
\begin{align}
    \mathbb{E}[\bra{0} U^{\dagger} \mathcal{O}_a^2 U \ket{0}^2 ] &= \frac{1}{d(d+1)} (\Tr[\mathcal{O}_a^2]^2 + \Tr[\mathcal{O}_a^4]) \nonumber \\
    &= \frac{1}{d(d+1)} (d^2 + \Tr[\mathcal{O}_a^4]) .
\end{align}
The next term can be evaluated with the third moment:
\begin{align}
    \mathbb{E}[\bra{0} U^{\dagger} \mathcal{O}_a^2 U \ket{0} \bra{0} U^{\dagger} \mathcal{O}_a U \ket{0}^2 ] = \frac{d^2 + 2 \Tr[\mathcal{O}_a^4]}{d(d+1)(d+2)} .
\end{align}
The final term follows a similar calculation with the fourth moment:
\begin{align}
    \mathbb{E}[ \bra{0} U^{\dagger} \mathcal{O}_a U \ket{0}^4 ] = \frac{3 d^2 + 6 \Tr[\mathcal{O}_a^4]}{d(d+1)(d+2)(d+3)} .
\end{align}
Combining these results at leading order in $d$ gives
\begin{align}
    \mathbb{E} [(\sigma_a^2 - 1)^2]  =  - d^{-1} + O (d^{-2})  + \Tr[\mathcal{O}_a^4] ( d^{-2} + O (d^{-3}) ) .
\end{align}
In order to understand the scaling of this variance with system size, we need information about how $\Tr[\mathcal{O}_a^4]$ scales.
Because $\Tr[\mathcal{O}_a^2]/d = 1$, we know that the eigenvalues $\lambda_{an}$ of $\mathcal{O}_a$ satisfy $\sum_n \lambda_{an}^2 = d$, or, treating $\lambda_a$ as a vector, $||\lambda_a^2||_1 = d$.
By the triangle inequality, we know that the largest magnitude these eigenvalues can have is 
\begin{align}
    || \mathcal{O}_a ||_{\infty} \leq \sum_{\alpha} \left| o^{(a)}_{\alpha} \right| || L_{\alpha} ||_{\infty} = \sum_{\alpha} \left| o^{(a)}_{\alpha} \right| ,
\end{align}
where we have used the fact that Pauli strings $L_{\alpha}$ have infinity norm of 1.
Furthermore, the normalization of the eigenvectors $o^{(a)}_{\alpha}$ means that \edit{$\sum_{\alpha} \left| o^{(a)}_{\alpha} \right|^2 = 1$.
By Cauchy-Schwarz, 
\begin{align}
    \sum_{\alpha} \left| o^{(a)}_{\alpha} \right| \leq \sqrt{N_L} \sqrt{\sum_{\alpha} \left| o^{(a)}_{\alpha} \right|^2} = \sqrt{N_L}.
\end{align}
As such, we have
\begin{align}
    || \mathcal{O}_a ||_{\infty} \leq \sqrt{N_L} ,
    \label{eq:inf norm of op}
\end{align}
where $N_L$ is the number of basis operators and grows as $O(\poly(N))$ by locality.
Thus, we know that $|\lambda_{an}| \leq \sqrt{N_L}$.}
Given this information, we now consider the object $\Tr[\mathcal{O}_a^4] = \sum_n \lambda_{an}^4$.
\edit{It is easy to see that
\begin{align}
    \sum_n \lambda_{an}^4 \leq \sum_n N_L \lambda_{an}^2 = d N_L,
\end{align}
which is just H\"older's inequality in the form $||\lambda_a^4||_1 \leq ||\lambda_a^2||_{\infty} ||\lambda_a^2||_1$.
As such, 
\begin{align}
    \Tr[\mathcal{O}_a^4] \leq d N_L .
\end{align}
}As expected, the polynomial growth in the number of basis operators cannot overcome the exponential growth in the size of the Hilbert space.
\edit{Thus, we find
\begin{align}
    \mathbb{E} [(\sigma_a^2 - 1)^2]  \leq  \frac{N_L - 1}{d} + O \left( \frac{1}{d^2} \right) .
\end{align}
}

Although this result implies a tight clustering of the variance spectrum around 1 and provides a scaling prediction for $D_E^2$, we can also make stronger statements about the clustering that predict the behavior of $\Delta_M$.
Indeed, we can use concentration of measure bounds which apply to Haar random averages \cite{Mele_2024,Ledoux_2001}. 
In particular, Lemma 53 of \cite{Mele_2024} tells us that 
\begin{align}
    \mathrm{Prob}\left( \left| \bra{v} \mathcal{O}_a \ket{v} \right| \geq \varepsilon \right) &\leq 2 \exp \left( - \frac{d \varepsilon^2}{18 \pi^3 || \mathcal{O}_a ||_{\infty}^2 } \right) , \nonumber \\
    \mathrm{Prob}\left( \left| \bra{v} \mathcal{O}_a^2 \ket{v} - 1 \right| \geq \varepsilon \right) &\leq 2 \exp \left( - \frac{d \varepsilon^2}{18 \pi^3 || \mathcal{O}_a ||_{\infty}^4 } \right) ,
\end{align}
where $\ket{v}$ is a Haar random state.
We expect that treating $\ket{v}$ as a Haar random state is a good approximation for all $\mathcal{O}_a$ orthogonal to the Hamiltonian (and with negligible overlap with higher powers of $H$).
Now note that $\bra{v} \mathcal{O}_a \ket{v}^2 \leq \left| \bra{v} \mathcal{O}_a \ket{v} \right| || \mathcal{O}_a ||_{\infty} $, and since 
\begin{align}
    \mathrm{Prob}( X \geq \varepsilon) \leq \mathrm{Prob}( Y \geq \varepsilon), 
\end{align}
for any $X \leq Y$, we can write
\begin{align}
    \mathrm{Prob}\left( \bra{v} \mathcal{O}_a \ket{v}^2 \geq \varepsilon \right) 
    &\leq \mathrm{Prob}\left( \left| \bra{v} \mathcal{O}_a \ket{v} \right| || \mathcal{O}_a ||_{\infty} \geq \varepsilon \right)  \nonumber \\
    &\leq 2 \exp \left( - \frac{d \varepsilon^2}{18 \pi^3 || \mathcal{O}_a ||_{\infty}^4 } \right)  .
\end{align}
We can now use the fact that sub-Gaussian random variables have a norm $|| \cdot ||_{\psi_2}$ to create a bound on the variance of $\mathcal{O}_a$.
In particular, let $X_1 \equiv \bra{v} \mathcal{O}_a^2 \ket{v} - 1$ and $X_2 \equiv \bra{v} \mathcal{O}_a \ket{v}^2$.
Then Definition 2.5.6 and Proposition 2.5.2 of Ref.~\cite{Vershynin_2018} tell us that
\begin{align}
    || X_1 ||_{\psi_2} &\leq C_1 \sqrt{\frac{18 \pi^3 || \mathcal{O}_a ||_{\infty}^4 }{d}}, \nonumber \\
    || X_2 ||_{\psi_2} &\leq C_2 \sqrt{\frac{18 \pi^3 || \mathcal{O}_a ||_{\infty}^4 }{d}} ,
\end{align}
where we use $C_i$ to denote absolute constants.
Then we can use the triangle inequality to write
\begin{align}
    || X_1 - X_2 ||_{\psi_2} \leq || X_1 ||_{\psi_2} + || X_2 ||_{\psi_2} \leq C_3 \sqrt{\frac{18 \pi^3 || \mathcal{O}_a ||_{\infty}^4 }{d}}
\end{align}
where we note that $X_1 - X_2 = \bra{v} \mathcal{O}_a^2 \ket{v} - \bra{v} \mathcal{O}_a \ket{v}^2 - 1$ and quantifies the difference between the operator's variance $\sigma_a^2$ and 1.
Furthermore, Proposition 2.5.2 of Ref.~\cite{Vershynin_2018} tells us that 
\begin{align}
    \mathrm{Prob}\left( \left| X_1 - X_2 \right| \geq \varepsilon \right) &\leq 2 \exp \left( - \frac{ \varepsilon^2}{K_{12}^2 } \right) ,
\end{align}
where
\begin{align}
    K_{12} \leq C_4 \sqrt{\frac{18 \pi^3 || \mathcal{O}_a ||_{\infty}^4 }{d}} .
\end{align}
This finally implies that
\begin{align}
    \mathrm{Prob}\left( \left| \sigma_a^2 - 1 \right| \geq \varepsilon \right) &\leq 2 \exp \left( - \frac{C d \varepsilon^2}{|| \mathcal{O}_a ||_{\infty}^4 } \right) ,
\end{align}
where $C$ is some absolute constant.
We therefore see that the chance that any single operator $\mathcal{O}_{a\notin K}$ has variance differing from $1$ by more than $\varepsilon$ decreases at least as fast as $1/\exp(\exp(N)/\mathrm{poly}(N))$.
Using a union bound on all $N_L - |K|$ operators orthogonal to the zero-variance subspace, we see that the probability that \textit{any} variance is more than $\varepsilon$ away from 1 must decay like
\edit{
\begin{align}
    \mathrm{Prob}\left( \bigcup_{a \notin K} \left| \sigma_a^2 - 1 \right| \geq \varepsilon \right) &\leq 2 (N_L - |K|) \exp \left( - \frac{C d \varepsilon^2}{ N_L^2 } \right) ,
    \label{eq:levy bound on all ops}
\end{align}
}which scales like $\mathrm{poly}(N)/\exp(\exp(N)/\mathrm{poly}(N))$ in system size.
Note that even if different operators have different absolute constants $C$, the bound \eqref{eq:levy bound on all ops} still stands as long as the smallest of these absolute constants is used.
We re-emphasize that this derivation relied on the assumption that all operators $\mathcal{O}_{a\notin K}$ have matrix elements that behave as though $\ket{v}$ were Haar random.

\section{Higher Moments of the Hamiltonian}
\label{app:krylov overlaps}

\begin{figure}[h!]
    \centering
    \includegraphics[width=0.6
\linewidth]{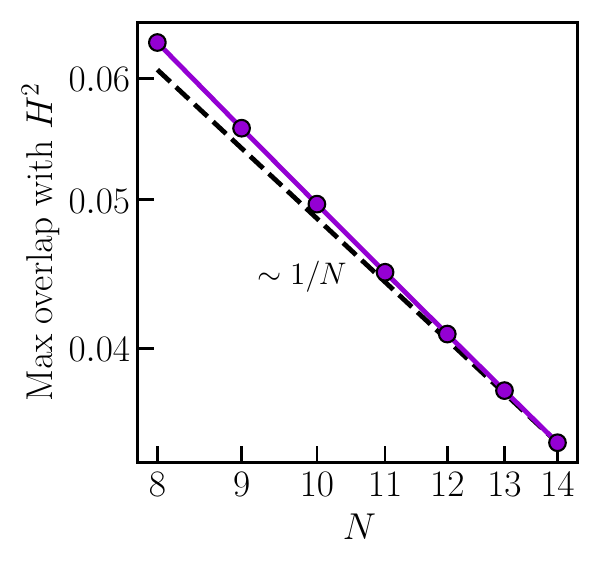}
    \caption{
    \textbf{Maximal Overlap with $H^2$.}
    To quantify how much an operator can deviate from the RMT predictions of Appendix~\ref{app:rmt predictions}, we numerically investigate the maximum overlap any \edit{geometrically} $2$-local operator can have [Eq.~\eqref{eq:max overlap with H2}] in the MFIM used in Fig.~\ref{fig:vars vs energy}.
    We observe the expected $1/N$ decay (linear on the log-log plot) with system size and furthermore verify that this overlap is not large enough to affect the observed gaps in Fig.~\ref{fig:vars vs energy}(e).
    }
    \label{fig:H2 overlaps}
\end{figure}

Although the covariance matrix $M$ automatically ensures that all operators $\mathcal{O}_{a\notin K}$ are Hilbert-Schmidt orthogonal to the Hamiltonian and other conserved charges, these operators may still have nonzero overlap with e.g. higher powers of $H$.
Consider an operator $\mathcal{O} = a P + b H^2$ decomposed into a linear combination of $H^2$ and an operator $P$ orthogonal to it.
The variance of this operator is then
\begin{align}
    \Var(\mathcal{O}) &= a^2 \Var(P) + b^2 \Var(H^2) + 2 a b \mathrm{Cov}(P H^2) \nonumber \\
    &= a^2 \Var(P).
\end{align}
The $b^2$ and $2ab$ terms vanish because $\ket{v}$ is an eigenstate of $H$.
Even if $\Var(P) \to 1$ in the ergodic limit, $\Var(\mathcal{O})$ will not since $|a| \leq 1$. 
Rather, the variance is determined by the overlap with $H^2$:
\begin{align}
    \Var(\mathcal{O}) \to a^2 = 1 - b^2 \frac{\Tr[H^4]}{d}
\end{align}
by the normalization of $\mathcal{O}$. Because $b = \Tr[\mathcal{O}H^2]/\Tr[H^4]$, this gives a prediction in the ergodic limit of 
\begin{align}
    \Var(\mathcal{O}) \to 1 - \frac{\Tr[\mathcal{O}H^2]^2}{d \Tr[H^4]}
     = 1 - \frac{(O|H^2)^2}{(H^2|H^2)}
     .
\end{align}
A similar calculation can be done for $H^3$ and higher powers provided that one accounts for the fact that powers of $H$ need not be orthogonal to each other.

Given this ability to deviate from RMT predictions, we would like to know how the maximal overlap of any $\mathcal{O}_a$ scales with system size $N$.
To estimate this, consider a translation-invariant Hamiltonian defined by $H = \sum_{i \mu} c_{i\mu} L_{i \mu}$, where $L_{i\mu}$ is an $\ell$-local Pauli centered on site $i$ and labeled by index $\mu$.
The normalization of $H$ tells us that the coefficients $c_{i\mu}$ decay like $1/\sqrt{N}$ with system size.
We can write
\begin{align}
    H^2 = \sum_{i\mu j\nu} c_{i\mu} c_{j\nu} \tilde{L}_{i\mu j\nu}
\end{align}
where $\tilde{L}_{i\mu j\nu}$ label a set of $O(N^2)$ \edit{geometrically} nonlocal Paulis.
Because these Paulis are still orthonormal, we expect $(H^2|H^2) \sim \mathrm{const}$ and $(L_{i\mu}|H^2) \sim 1/N$.
Although this implies that a single basis operator has overlap $\sim 1/N^2$ with $H^2$, the maximum overlap that any local operator can have with $H^2$ will get contributions from all $O(N)$ basis operators.
As such, we expect the maximal overlap of any local operator with $H^2$ to scale as $1/N$.
In Fig.~\ref{fig:H2 overlaps}, we confirm this scaling at small system sizes in the mixed field Ising model from Fig.~\ref{fig:vars vs energy} by plotting
\begin{align}
    \sum_{\alpha} \frac{(L_{\alpha}|H^2)^2}{(H^2|H^2)} .
    \label{eq:max overlap with H2}
\end{align}

A similar analysis can be done for higher powers of $H$, but this already poses an important exception to the predictions of Appendix~\ref{app:rmt predictions}.
Indeed, as $N$ grows, a single operator should emerge from the variance spectrum with a variance scaling like $1 - C/N$ for some constant $C$ while the rest of the spectrum continues to exponentially concentrate around 1.
This operator will maximize its overlap with $H^2$ to minimize its variance, and eventually other operators will do the same with the component of $H^3$ orthogonal to $H^2$ and $H$, and so on in a Gram-Schmidt orthogonality construction.

One can make similar arguments to that above about how these overlaps scale as inverse polynomials of higher and higher order ($1/N^2$,$1/N^3$, etc.).
As such, they represent an important alteration of the predicted variance spectrum compared to that of a truly Haar random state.
Although these corrections are not large enough to affect our numerics at current system sizes, future work could investigate how to modify the proposed metrics to account for energy conservation, similar to how Ref.~\cite{Rodriguez-Nieva_2024} did for the distribution of entanglement entropy.

\section{Free Fermion Calculations for Variance Spectra}
\label{app:ff predictions}

\begin{figure*}[t]
    \centering
    \includegraphics[width=\linewidth]{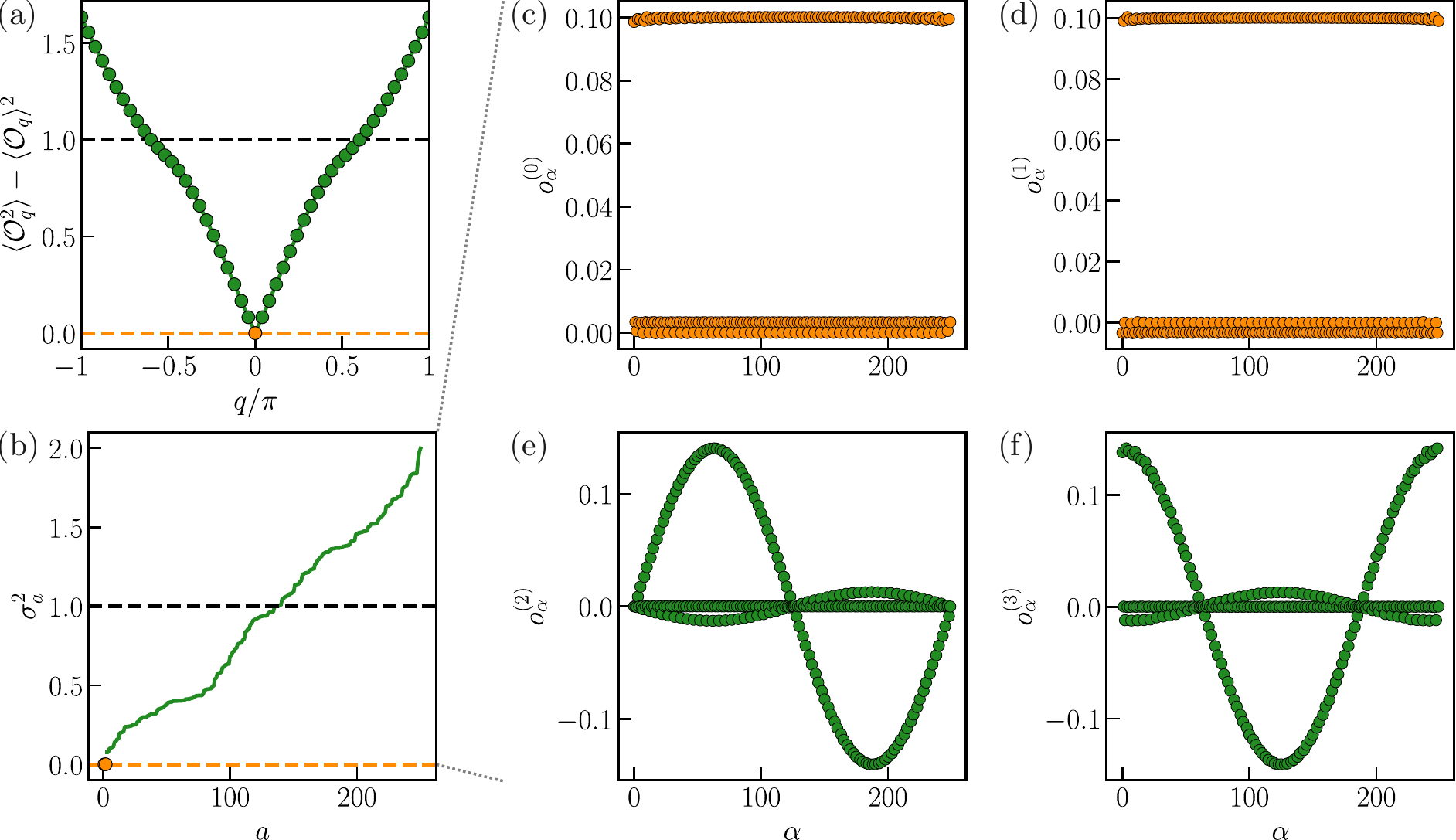}
    \caption{
    \textbf{Hamiltonian Learning in Free Fermion Systems.}
    We consider a periodic $N=50$ TFIM with no boundary fields as a numerical testbed for our analytic arguments for the anti-concentration of the variance spectrum in free fermion systems.
    (a) We first calculate the variance of our cosine-deformed Hamiltonian [Eq.~\eqref{eq:cosine deformed H}] in a state at half filling (although not a generic high-temperature eigenstate, it most cleanly shows the properties we use to construct our bounds) as a function of momentum $q$.
    As expected, the variance spectrum is gapless as we tune towards the $q=0$ Hamiltonian.
    (b) We then construct the covariance matrix in a free-fermion basis [Eq.~\eqref{eq:ff basis}] and plot the resulting variance spectrum.
    The first (c) and second (d) eigenvectors (with zero variance) correspond to the Hamiltonian and the other 2-local integral of motion in the TFIM.
    In contrast, the third (e) and fourth (f) eigenvectors correspond to exactly the cosine (and sine) deformations used to construct our anti-concentration bound.
    }
    \label{fig:ff numerics}
\end{figure*}

We will now show how the simple and exact solutions available in translation-invariant free fermion models allow for an analytic understanding of the small gaps and large spread of the variance spectrum.
We follow some of the conventions used in \cite{Dziarmaga_2005}.
In general, we will take the Hamiltonian to have the form
\begin{align}
    H = \sum_{i=1}^{N} h_i ,
\end{align}
with periodic boundary conditions.
The operators $h_i$ are geometrically $\ell$-local (we will use $k$ to label wavenumber).
Because this model is translationally invariant and can be mapped to free fermions, we can always rewrite it as
\begin{align}
    H = \sum_k \begin{pmatrix}
        c_k^{\dagger} & c_{-k}
    \end{pmatrix}
    H(k)
    \begin{pmatrix}
        c_k \\
        c_{-k}^{\dagger}
    \end{pmatrix}
    + \mathrm{const} ,
    \label{eq:ff form}
\end{align}
where 
\begin{align}
    H(k) = \begin{pmatrix}
        H_{00}(k) & H_{01}(k) \\
        H_{01}^*(k) & -H_{00}(k)
    \end{pmatrix} ,
\end{align}
and we have enforced hermiticity. The additive constant allows us to make $H(k)$ traceless as well.
This structure tells us that there exists a unitary $U(k)$ which diagonalizes $H(k)$ such that 
\begin{align}
    H(k) = U(k) \Lambda(k) U^{\dagger}(k), \quad \Lambda(k) = \frac{1}{2} \begin{pmatrix}
        \epsilon(k) & 0 \\
        0 & -\epsilon(k)
    \end{pmatrix} .
\end{align}
This transformation turns the full Hamiltonian into
\begin{align}
    H = \sum_k \begin{pmatrix}
        \gamma_k^{\dagger} & \gamma_{-k}
    \end{pmatrix}
    \Lambda(k)
    \begin{pmatrix}
        \gamma_k \\
        \gamma_{-k}^{\dagger}
    \end{pmatrix}
    = \sum_k \epsilon(k) \left( \gamma^{\dagger}_k \gamma_k - \frac{1}{2} \right) , 
\end{align}
where we assume $\epsilon(-k) = \epsilon(k)$.
We will also assume $H(k)$, $U(k)$, and $\epsilon(k)$ are all smooth functions of $k$.
Every eigenstate of this system can now be written as a list of occupations of each $k$ mode.

To understand how this can lead to small gaps in the variance spectrum, consider the following operator constructed by taking the Hamiltonian and applying a cosine envelope:
\begin{align}
    \mathcal{O}_q = \sum_{i=1}^{N} \cos (qi) h_i ,
    \label{eq:cosine deformed H}
\end{align}
where the lattice spacing is set to $a = 1$.
Note that this operator is orthogonal to $H$ by construction whenever $q = 2 \pi n / N$ and integer $n \neq 0$.
This construction also works if we use a sine envelope, and furthermore also works if we apply either envelope to a local integral of motion $Q$ of the Hamiltonian.
Because this operator now has two components with momentum $\pm q$, it will be written in the following form in momentum space:
\begin{align}
    \mathcal{O}_q &= \frac{1}{2} \sum_k \begin{pmatrix}
        c_{k+q}^{\dagger} & c_{-k-q}
    \end{pmatrix}
    H(k)
    \begin{pmatrix}
        c_k \\
        c_{-k}^{\dagger}
    \end{pmatrix} \nonumber \\
    &+ \begin{pmatrix}
        c_{k-q}^{\dagger} & c_{-k+q}
    \end{pmatrix}
    H(k)
    \begin{pmatrix}
        c_k \\
        c_{-k}^{\dagger}
    \end{pmatrix}
    + O(q) .
\end{align}
Although the Fourier transform can modify $H(k)$ due to factors like $e^{iq(\ell-1)}$, the difference is clearly $O(q)$, and as such, we drop the term.
We would like to compute the variance of this operator to order $O(q^0)$. 
If it is zero, then tuning $q$ will allow us to continuously increase the variance of $\mathcal{O}$, showing that the variance spectrum is gapless.
Now, let us transform to the diagonal fermion basis:
\begin{align}
    \mathcal{O}_q &= \frac{1}{2} \sum_k \begin{pmatrix}
        \gamma_{k+q}^{\dagger} & \gamma_{-k-q}
    \end{pmatrix}
    \Lambda(k)
    \begin{pmatrix}
        \gamma_k \\
        \gamma_{-k}^{\dagger}
    \end{pmatrix} \nonumber \\
    &+ \begin{pmatrix}
        \gamma_{k-q}^{\dagger} & \gamma_{-k+q}
    \end{pmatrix}
    \Lambda(k)
    \begin{pmatrix}
        \gamma_k \\
        \gamma_{-k}^{\dagger}
    \end{pmatrix}
    + O(q) .
\end{align}
Although $U(k\pm q)$ is required to diagonalize the row vectors, smoothness ensures that $U(k\pm q) = U(k) + O(q)$.
This finally gives
\begin{align}
    \mathcal{O}_q = \frac{1}{2} \sum_k \epsilon(k) \left( \gamma^{\dagger}_{k+q} \gamma_k + \gamma^{\dagger}_{k-q} \gamma_k 
    % - 1 
    \right)
    + O(q) .
\end{align}
It is clear that $\langle \mathcal{O}_q \rangle = 0$ for any $q \neq 0$ when calculated with respect to any eigenstate of $H$.
In contrast, the variance has contributions given by
\begin{align}
    &\langle \mathcal{O}_q^2 \rangle - \langle \mathcal{O}_q \rangle^2 = \langle \mathcal{O}_q^2 \rangle \nonumber \\
    &= \frac{1}{4} \sum_k \epsilon^2(k) (\langle \gamma^{\dagger}_{k} \gamma_{k+q} \gamma^{\dagger}_{k+q} \gamma_k \rangle + \langle \gamma^{\dagger}_{k} \gamma_{k-q} \gamma^{\dagger}_{k-q} \gamma_k \rangle)
    + O(q) \nonumber \\
    &= \frac{1}{4} \sum_k \epsilon^2(k) ( n_k (1 - n_{k+q}) + n_k (1 - n_{k-q}) )
    + O(q) ,
\end{align}
where $n_k = \langle \gamma_k^{\dagger} \gamma_k \rangle$ is the occupation of the $k$th mode and we have used the fact that $\epsilon(k\pm q) = \epsilon(k) + O(q)$.

It is therefore clear that all variance is generated when an occupied and unoccupied mode differ in momentum by $q$.
As such, we can immediately see that variance is generated at ``Fermi surfaces'' where the mode occupation changes from 1 to 0 (or vice versa).
For the ground state (and highest excited state), there are no Fermi surfaces, so the variance is gapless.
Furthermore, a state with modes filled up consecutively until some Fermi momentum $k_F$ will have two points in its Fermi surface ($\pm k_F$).
Each of these points will contribute once to the variance, making the total $\epsilon^2(k_F)/2$.
However, we still need to normalize the operator to understand whether this is large or small.
To do so, note that $\Tr(A^2)/d$ for some operator $A$ is equivalent to averaging $\bra{v} A^2 \ket{v}$ over all eigenstates $v$ of our Hamiltonian.
As such, we consider 
\begin{align}
    \mathbb{E}_v \bra{v} \mathcal{O}_q^2 \ket{v} &= \frac{1}{4} \sum_k \epsilon^2(k) \left( \frac{1}{2} \cdot \frac{1}{2} + \frac{1}{2} \cdot \frac{1}{2} \right)
    + O(q) \nonumber \\
    &= \frac{1}{8} \sum_k \epsilon^2(k) + O(q) ,
\end{align}
since each mode is on average half full.
Thus, the normalized variance is
\begin{align}
    \sigma_q^2 = \frac{2 \sum_k \epsilon^2(k) ( n_k (1 - n_{k+q}) + n_k (1 - n_{k-q}) )}{\sum_k \epsilon^2(k)} + O(q) .
\end{align}
Note that $\epsilon(k) = O(1)$ in system size, so the denominator scales as $O(N)$ while the numerator is constant for a constant number of Fermi surfaces. As such, the variance spectrum is also gapless in the thermodynamic limit.
In fact, \textit{any} state where the numerator grows as $o(N)$ will have gapless variance in the thermodynamic limit.
Intuitively, this suggests that the number of Fermi surfaces should grow more slowly than $N$.
However, even this is not the full story, as the state with the maximum number of Fermi surfaces, where the occupation list looks like $0,1,0,1,\dots$ will still have gapless variance.
We just need to choose $q$ to connect every \textit{other} mode, such that $q = 2 q_0$, where $q_0 = 2\pi/N$ is the smallest allowed momentum for a fixed system size.
If we choose this value for $q$, $\mathcal{O}_q$ will never connect an occupied and unoccupied mode.

It is therefore clear that there are many states in the middle of the Hamiltonian spectrum with gapless variance.
However, for a random set of mode occupations, the operator normalization requires that the variance is (on average) 1.
This does not automatically rule out gaplessness, as we have only considered a single ansatz for the lowest variance operator that is not the Hamiltonian.
Furthermore, the number of states in the spectrum that have a \textit{small but finite} gap is \textit{exponential} in system size, and this fact allows us to construct an anti-concentration bound on the variance spectrum.

Specifically, consider the operator $\mathcal{O}_{q_0}$.
The variance of this operator (to leading order) gives an upper bound on the gap $\Delta_M$ of the entire variance spectrum:
\begin{align}
    \Delta_M \leq \frac{2}{Z} \sum_k \epsilon^2(k) [ n_k (1 - n_{k+q_0}) + n_k (1 - n_{k-q_0}) ]
\end{align}
where $Z = \sum_k \epsilon^2(k)$.
Let us now make a ``duality'' map in the same sense of the Kramers-Wannier map of spin configurations in the Ising model.
For any binary list of mode occupations $n_k$, we will define the dual list of ``Fermi surface occupations'' $\Tilde{n}_k \equiv n_k (1 - n_{k+q_0}) + n_{k+q_0} (1 - n_k)$.
This value is $0$ if $n_k = n_{k+q_0}$, i.e. the occupations are the same and there is no Fermi surface, and $1$ otherwise.
Let us now upper bound the gap for a given state.
We will use $\epsilon_{\rm max}$ to indicate the maximum of $|\epsilon(k)|$ over the entire Brillouin zone.
Then 
\begin{align}
    \Delta_M &\leq \frac{2 \epsilon^2_{\rm max}}{Z}  \sum_k   n_k (1 - n_{k+q_0}) + n_{k+q_0} (1 - n_{k}) \nonumber \\
    &= \frac{2 \epsilon^2_{\rm max}}{Z}  \sum_k \Tilde{n}_k .
\end{align}
As such, the gap is upper bounded by an expression proportional to the number of Fermi surfaces.
For a given state, let us denote this quantity by $N_{\rm FS} = \sum_k \Tilde{n}_k$ (which must be even by the periodicity of the Brillouin zone).
For large $N$, we can rewrite this as 
\begin{align}
    \Delta_M \leq \frac{2 \epsilon^2_{\rm max}}{z}  \frac{N_{\rm FS}}{N} = \frac{2 \epsilon^2_{\rm max}}{z} p
\end{align}
where $z = Z/N$ (well-defined in the thermodynamic limit by an integral) and $p = N_{\rm FS}/N$ is the density of Fermi surfaces in the Brillouin zone.
The number of states in the spectrum with this number of Fermi surfaces is clearly given by
\begin{widetext}
\begin{align}
    2 \begin{pmatrix}
        N \\
        N_{\rm FS}
    \end{pmatrix}
    &= \frac{2 N!}{(pN)!((1-p)N)!} \nonumber \\
    &\sim \frac{2 \sqrt{2 \pi N} N^N e^{-N}}{\sqrt{2 \pi p N} (pN)^{pN} e^{-pN} \sqrt{2 \pi N (1-p)} ((1-p)N)^{(1-p)N} e^{-(1-p)N}} \nonumber \\
    &= \frac{2}{\sqrt{2 \pi p (1-p) N} p^{pN} (1-p)^{(1-p)N}} \nonumber \\
    &= \frac{2^{N S_b(p) + 1}}{\sqrt{2 \pi p (1-p) N}} 
\end{align}
\end{widetext}
by Stirling's approximation, where
\begin{align}
    S_b(p) = - p \log_2 p - (1-p) \log_2 (1-p)
\end{align}
is the binary entropy function.
The 2 follows from the fact that the map $n_k \to 1-n_k$ preserves the numbers $\tilde{n}_k$.
For a random free fermion state, the probability that the gap is at most $\Delta_p = 2 \epsilon^2_{\rm max} p/z$ is therefore
\begin{align}
    \mathrm{Prob}\left( \Delta_M \leq \Delta_p \right) \geq \frac{2}{d} \begin{pmatrix}
        N \\
        p N
    \end{pmatrix} 
    \approx \frac{2 d^{S_b(p) - 1}}{\sqrt{2 \pi p (1-p) N}}
\end{align}
after Stirling's approximation is used.
Although this bound does decay exponentially, we note that it decays much more slowly than the L\'evy bound found for Haar random states in the ergodic regime.
In fact, we can convert this asymptotic bound to a similar form as Eq.~\eqref{eq:levy bound for var spectrum}:
\begin{align}
    \mathrm{Prob}\left( \bigcup_{a\notin K} \left| \sigma_a^2 - 1 \right| 
    \geq \varepsilon \right)
    &\geq \mathrm{Prob}\left( \bigcup_{a\notin K} \sigma_a^2  
    \leq 1 - \varepsilon \right) \\
    % &\geq \mathrm{Prob}\left( 1 - \Delta_M \geq \varepsilon \right) \nonumber \\
    &=  \mathrm{Prob}\left( \Delta_M \leq 1 - \varepsilon \right)  \nonumber \\
    &\geq \frac{2}{d} \begin{pmatrix}
        N \\
        p(\varepsilon) N
    \end{pmatrix}  \nonumber \\
    &\approx \frac{2 d^{S_b(p(\varepsilon)) - 1}}{\sqrt{2 \pi p(\varepsilon) (1-p(\varepsilon)) N}}
\end{align}
where we have converted $\varepsilon$ into a Fermi surface density using
\begin{align}
    p(\varepsilon) = \frac{z}{2 \epsilon_{\rm max}^2} (1 - \varepsilon) .
    \label{eq:p function definition}
\end{align}
This shows that in the thermodynamic limit, the probability that any variance in the spectrum (besides the Hamiltonian's) differs from $1$ by at least $\varepsilon$ is \textit{lower bounded} by a function that scales like $1/(\exp(N)\sqrt{N})$, a result of simply counting states for which our low-variance ansatz works.
Furthermore, the exponential grows more slowly than $d$.
This result stands in stark contrast to the concentration bound we found for Haar random states and shows a fundamental difference in the concentration of the variance spectrum between free fermion and ergodic systems.

\vspace{0.2cm}

Before moving to numerical demonstrations of these low-variance operators in large free fermion systems, we consider another metric of interest: how much the normalized variance fluctuates around the mean value of $1$.
This requires us to calculate
\begin{widetext}
\begin{align}
    \mathbb{E}_v [\sigma_q^4] = \frac{4}{Z^2} \sum_{kp} \epsilon^2(k) \epsilon^2(p) &\Bigg\{
    \mathbb{E}_v [n_k (1 - n_{k+q}) n_p (1 - n_{p+q})] 
    + \mathbb{E}_v [n_k (1 - n_{k+q}) n_p (1 - n_{p-q})] \nonumber \\
    &+ \mathbb{E}_v [n_k (1 - n_{k-q}) n_p (1 - n_{p+q})] 
    + \mathbb{E}_v [n_k (1 - n_{k-q}) n_p (1 - n_{p-q})] 
    \Bigg\}
    + O(q) .
\end{align}
\end{widetext}
The leading contribution to this expression (which scales as $O(1)$ in system size) is given by the case when none of the $k$ values in any expression are the same.
Thus, the averages are all uncorrelated and can be taken independently.
Every term becomes $1/2^4$, and the expected value is $1$.
However, as we care about the fluctuations, this simply cancels the offset $\mathbb{E}_v [\sigma_q^2]^2 = 1$.
Thus, the first term that survives scales as $1/N$ in system size and is proportional to
\begin{align}
    \mathbb{E}_v [\sigma_q^4] - 1 \propto \frac{4}{Z^2} \sum_k \epsilon^4(k) + O(q) = O(1/N),
\end{align}
in agreement with the numerics in Fig.~\ref{fig:vars vs energy}(d) and previous work \cite{Alba_2015,Essler_2024,LeBlond_2019}.

Given these analytic arguments, we now turn to numerical simulations of an $N = 50$ TFIM with periodic boundary conditions and no boundary fields ($h_1 = h_N = 0$).
We can access such system sizes exactly by explicitly converting to the free fermion form \eqref{eq:ff form} and diagonalizing.
In Fig.~\ref{fig:ff numerics}, we consider the eigenstate of this model corresponding to half filling.
Although this state is a highly structured state and not representative of midspectrum eigenstates, it gives the cleanest example of our above results.
We find similar results if one considers states with e.g. more Fermi surfaces or staggered occupations (as discussed above), but the gaps become larger and the cosine-deformed operators may rise slightly in the variance spectrum.

Given these considerations, we first calculate the variance of our cosine-deformed Hamiltonian $\mathcal{O}_q$ at half-filling as a function of $q$.
In Fig.~\ref{fig:ff numerics}(a), we see that as $q$ is tuned away from 0, the variance continuously increases from 0 (since $\mathcal{O}_{q=0} = H$ has zero variance).
We choose $q = 2\pi n/N$ such that all operators with $n \neq 0$ are orthogonal to the Hamiltonian.
As expected, our cosine-deformed Hamiltonian is gapless and therefore implies that the variance spectrum is also gapless as $N \to \infty$.

However, this operator does not simply provide a good upper bound on the variance spectrum's gap; in many cases, it is explicitly found as the operator with the lowest nonzero variance in the spectrum of $M$.
This feature is nicely seen in the remaining subfigures of Fig.~\ref{fig:ff numerics}.
First, we construct a basis of operators to calculate $M$ using just free fermion operators.
For each site $j$, there are 5 basis operators:
\begin{align}
    &X_j, Z_j Z_{j+1}, Z_j Y_{j+1}, Y_j Z_{j+1}, Y_j Y_{j+1} \nonumber \\
    \sim \: & c^{\dagger}_j c_j, c^{\dagger}_j c_{j+1}, c^{\dagger}_j c^{\dagger}_{j+1},
    c_j c^{\dagger}_{j+1},
    c_j c_{j+1}
    \label{eq:ff basis}
\end{align}
where we give a decomposition in both the spin basis and the fermion basis (note that the basis operators do not directly correspond to one another after a Jordan-Wigner transformation; the fermion bilinears are some linear combination of the transformed spin operators and vice versa).
Given this basis, we construct $M$ for our half-filled state and plot the resulting variance spectrum in Fig.~\ref{fig:ff numerics}(b).
This spectrum clearly has a small gap and a large spread.
Both 2-local (translationally invariant) integrals of motion in the TFIM ($H$ and $Q_2$ \cite{Grady_1982,Prosen_1998,fagotti2013reduced,essler2016quench,vidmar2016generalized}) are learned in the zero variance subspace, and we plot the corresponding eigenvectors $o_{\alpha}^{(a=0)}$ and $o_{\alpha}^{(a=1)}$ of $M$ in Fig.~\ref{fig:ff numerics}(c-d).
In contrast, the next two eigenvectors (which have nonzero variance) clearly break translation invariance in exactly the same way our cosine-deformed Hamiltonian does.
Aside from small fluctuations, 
these operators clearly follow a sinusoid profile.
As such, our ansatz in Fig.~\ref{fig:ff numerics}(a) is not just a bound but also partially describes the low-variance spectrum of $M$.

\section{Connecting the Variance Spectrum to Other Metrics}
\label{app:variance spectrum connections}

In this appendix, we elucidate the connections between the variance spectrum and other ergodicity metrics and learning algorithms.
Indeed, it has been shown that methods similar to this approach are able to identify conserved and nearly conserved quantities \cite{Bentsen_2019,Zhan_2024} in many different settings.
Treating conserved operators as the ``ground states'' of some more general superoperator also extends beyond the covariance matrix and has proven useful in characterizing integrals of motion in general quantum systems \cite{Moudgalya_2023_num,Moudgalya_2024,Li_2025,Pawlowski_2025}.

To begin, we will show that the matrix elements of $M$ are sufficient to calculate the purity (and therefore R\'enyi-2 entanglement entropy) of any \edit{geometrically} $\ell$-local subsystem.
Let $\Lambda$ denote the set of $4^{\ell}-1$ nontrivial Pauli strings on a contiguous subsystem of $\ell$ qubits.
This set defines a submatrix of $M$, the normalized trace of which is
\begin{align}
    2^{-\ell} \Tr M_{\Lambda} = 2^{-\ell} \sum_{\alpha \in \Lambda} M_{\alpha \alpha} = 2^{-\ell} \sum_{\alpha \in \Lambda} (1 - \langle L_{\alpha}\rangle^2) .
\end{align}
Focusing on the latter term in the summand, we have
\begin{align}
    \sum_{\alpha \in \Lambda} \langle L_{\alpha}\rangle^2 
    = 2^{2\ell} \sum_{\alpha \in \Lambda} (\rho_{\Lambda}|L_{\alpha}) (L_{\alpha}|\rho_{\Lambda})
\end{align}
using $(A|B) \equiv \Tr(A^{\dagger}B)/2^{\ell}$ as a rescaled Hilbert-Schmidt inner product and $\rho_{\Lambda}$ as the reduced density matrix on the subsystem of $\ell$ qubits.
This expression looks almost like a resolution of the identity has been inserted into the inner product $(\rho_{\Lambda}|\rho_{\Lambda})$.
To make this true, we need to include the identity operator $I$ to create a complete operator basis.
Because $(I|\rho_{\Lambda}) = 2^{-\ell}$ by normalization, this is easily accomplished:
\begin{align}
    2^{2\ell} \sum_{\alpha \in \Lambda} (\rho_{\Lambda}|L_{\alpha}) (L_{\alpha}|\rho_{\Lambda})
    &= 2^{2\ell} \left[ (\rho_{\Lambda}|\rho_{\Lambda}) - 2^{-2\ell} \right] \nonumber \\
    &= 2^{\ell} \Tr(\rho_{\Lambda}^2) - 1
\end{align}
Thus, the original trace is equivalent to
\begin{align}
    2^{-\ell} \Tr M_{\Lambda} &= 2^{-\ell} ( 4^{\ell} - 1 - 2^{\ell} \Tr(\rho_{\Lambda}^2) + 1) \nonumber \\
    &= 2^{\ell} - \Tr(\rho_{\Lambda}^2).
\end{align}
As such, the variance spectrum is sufficient to extract the purity of any contiguous $\ell$-qubit subsystem.
Although this normalized trace has an exponential constant prefactor $2^{\ell}$, this value does not grow with system size $N$ and is furthermore only $2^2 = 4$ in the models we consider.
However, it is also clear that only considering purities discards useful information about the spread of the variance spectrum, so we opt to consider the variance spectrum as a whole for our metrics.

The covariance matrix also represents the tomographically-complete limit of the learning algorithm in Ref.~\cite{Lindner-Bairey_2019}.
The algorithm uses a rectangular matrix $K_{\alpha \beta}$ constructed from two operator bases $A_{\alpha}$ and $L_{\alpha}$ to learn the Hamiltonian:
\begin{align}
    K_{\alpha \beta} = \langle i [A_{\alpha},L_{\beta}] \rangle.
\end{align}
The $L_{\alpha}$ operators still constitute the basis from which the Hamiltonian is built, whereas we will refer to the $A_{\alpha}$ as constraint operators.
Instead of requiring that the Hamiltonian have zero variance on $\ket{v}$ as in \cite{Qi_2019}, this method requires that the expectation value of any operator $\bra{v} A_{\alpha} \ket{v}$ be constant under evolution by $H$:
\begin{align}
    \partial_t \langle A_{\alpha} \rangle = - \langle i [A_{\alpha},H] \rangle = 0.
\end{align}
As such, the $A_{\alpha}$ operators are used to impose constraints on the possible form of $H$.

It was already shown in Appendix E of Ref.~\cite{Lindner-Bairey_2019} that when all $4^N$ possible constraints $A_{\alpha}$ are imposed, the product $K^T K$ reduces to the covariance matrix $M$.
Here, we will show that this object more generally corresponds to a redefinition of QFI when the constraints only span the basis of a subsystem $\Lambda$.
First, note that we can define
\begin{align}
    \rho(\theta_{\alpha}) = e^{- i \theta_{\alpha} L_{\alpha}} \rho e^{i \theta_{\alpha} L_{\alpha}}
\end{align}
where $\rho = \ket{v}\bra{v}$ is the system's density matrix and $\theta_{\alpha}$ is the angle by which we rotate this state using the generator $L_{\alpha}$.
Then $\partial_{\alpha} \rho = - i [L_{\alpha},\rho]$.
Using this, we can write
\begin{align}
    (K^T K)_{\alpha \beta} 
    &= \sum_{\gamma} K_{\gamma \alpha} K_{\gamma \beta} \nonumber \\
    &= \sum_{\gamma} i^2 \Tr(\rho [A_{\gamma},L_{\alpha}]) \Tr(\rho [A_{\gamma},L_{\beta}]) \nonumber \\
    &= \sum_{\gamma} (- i)^2\Tr(A_{\gamma}[L_{\alpha},\rho]) \Tr( A_{\gamma} [L_{\beta},\rho]) \nonumber \\
    &= \sum_{\gamma} \Tr(A_{\gamma} \partial_{\alpha} \rho) \Tr( A_{\gamma} \partial_{\beta} \rho)
\end{align}
Let us now assume that the $A_{\gamma}$ span all basis operators over a subsystem denoted by $S$ with $|S|$ qubits.
As such, we can trace out the rest of the system $\overline{S}$ to get
\begin{align}
    \sum_{\gamma} \Tr(A_{\gamma} \partial_{\alpha} \rho_S) \Tr( A_{\gamma} \partial_{\beta} \rho_S)
\end{align}
where we have used that $\Tr_{\overline{S}} (\partial_{\alpha}\rho) = \partial_{\alpha} \Tr_{\overline{S}} (\rho) = \partial_{\alpha} \rho_S$ by linearity.
Now, we can define $(A|B) = \Tr(A^{\dagger}B)/2^{|S|}$ to rewrite this as
\begin{align}
    2^{2|S|} \sum_{\gamma} (\partial_{\alpha} \rho_S | A_{\gamma}) (A_{\gamma} | \partial_{\beta} \rho_S)
    = 2^{|S|} \Tr(\partial_{\alpha} \rho_S \partial_{\beta} \rho_S).
\end{align}
If $S$ is the entire system such that $\rho_S = \rho$ is a pure state, one can show that this reduces to the rescaled QFI matrix $d M_{\alpha \beta}$ \cite{Qi_2019,Lindner-Bairey_2019}.
If, however, we consider a subsystem with mixed state $\rho_S$, the resulting matrix corresponds to a modification of the standard QFI.
Recall that the standard definition of quantum Fisher information follows from calculating $\partial_{\alpha}^2 D_Q(\rho(\theta_{\alpha})||\rho(0)) |_{\theta_{\alpha}=0}$, where
\begin{align}
    D_Q(\rho(\theta_{\alpha})||\rho(0)) = \Tr(\rho(\theta_{\alpha}) \log \rho(\theta_{\alpha}) - \rho(\theta_{\alpha}) \log \rho(0))
\end{align}
is the quantum relative entropy, the quantum analog of the Kullback-Leibler divergence.
Although taking this derivative requires the use of the symmetric logarithmic derivative \cite{Liu_2019}, our modification will not.
Specifically, we consider the object 
\begin{align}
    D_2(\rho(\theta_{\alpha}),\rho(0)) = ||\rho(\theta_{\alpha}) - \rho(0)||_2^2 = \Tr[(\rho(\theta_{\alpha}) - \rho(0))^2] .
\end{align}
It is then easy to see that 
\begin{align}
    \partial_{\alpha} \partial_{\beta} D_2(\rho(\vec{\theta}),\rho(0)) \Big|_{\vec{\theta}=0} 
    = 2 \Tr(\partial_{\alpha} \rho \partial_{\beta} \rho)
\end{align}
for the multi-parameter case.
Up to a scale factor, this is equivalent to $K^T K$, and as such, the learning method in Ref.~\cite{Lindner-Bairey_2019} still probes the sensitivity of the state $\rho$ when the constraints are restricted to a subsystem.

\edit{
\section{Metric Sensitivity Details}
\label{app:metric sensitivity details}
}

\begin{figure}[h!]
    \centering
    \includegraphics[width=
\linewidth]{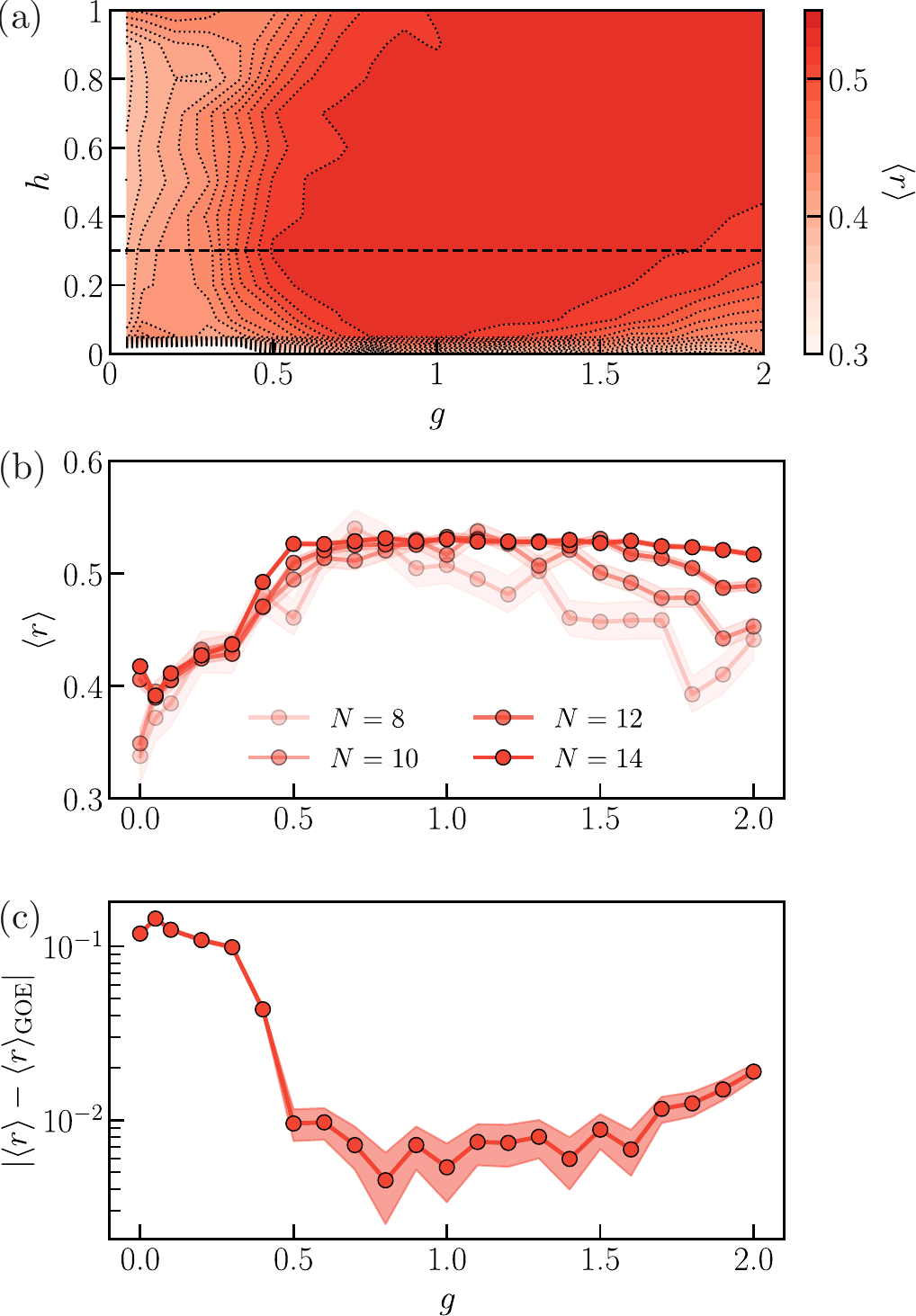}
    \caption{
    \edit{
    \textbf{Level Spacing Ratio with the Full Spectrum.}
    (a) We plot the level spacing ratio $\langle r \rangle$ averaged over the entire energy spectrum (rather than 600 eigenstates near $E=0$) for the parameters in Fig.~\ref{fig:ergodicity maps}.
    Even with the reduced noise from eigenstate sampling, this map does not show any substantial features within the ergodic regime.
    (b) We also consider the behavior of the metric along $h=0.3$ [black dashed line in (a)] for various system sizes, showing the standard error of the mean over the entire spectrum as ribbons.
    The substantial reduction in noise compared to Fig.~\ref{fig:snr cuts}(a) allows us to use this metric as a proxy for the ``signal'' in the SNR calculations in the main text.
    (c) However, the reduction in fluctuations still is not enough to resolve structure in the ergodic regime ($g \gtrsim 0.5$) for $N = 14$.
    Since we have no proxy for the signal in this case, we opt for plotting the absolute deviation of $\langle r \rangle$ from the prediction for the Gaussian Orthogonal Ensemble (GOE) $\langle r \rangle_{\rm GOE} \approx 0.536$ on a log scale \cite{Rodriguez-Nieva_2024,Atas_2013}.
    Even here, the error bars are too large to extract any meaningful signal in the ergodic regime. 
    }
    }
    \label{fig:full level spacing}
\end{figure}

\begin{figure*}[t!]
    \centering
    \includegraphics[width=\linewidth]{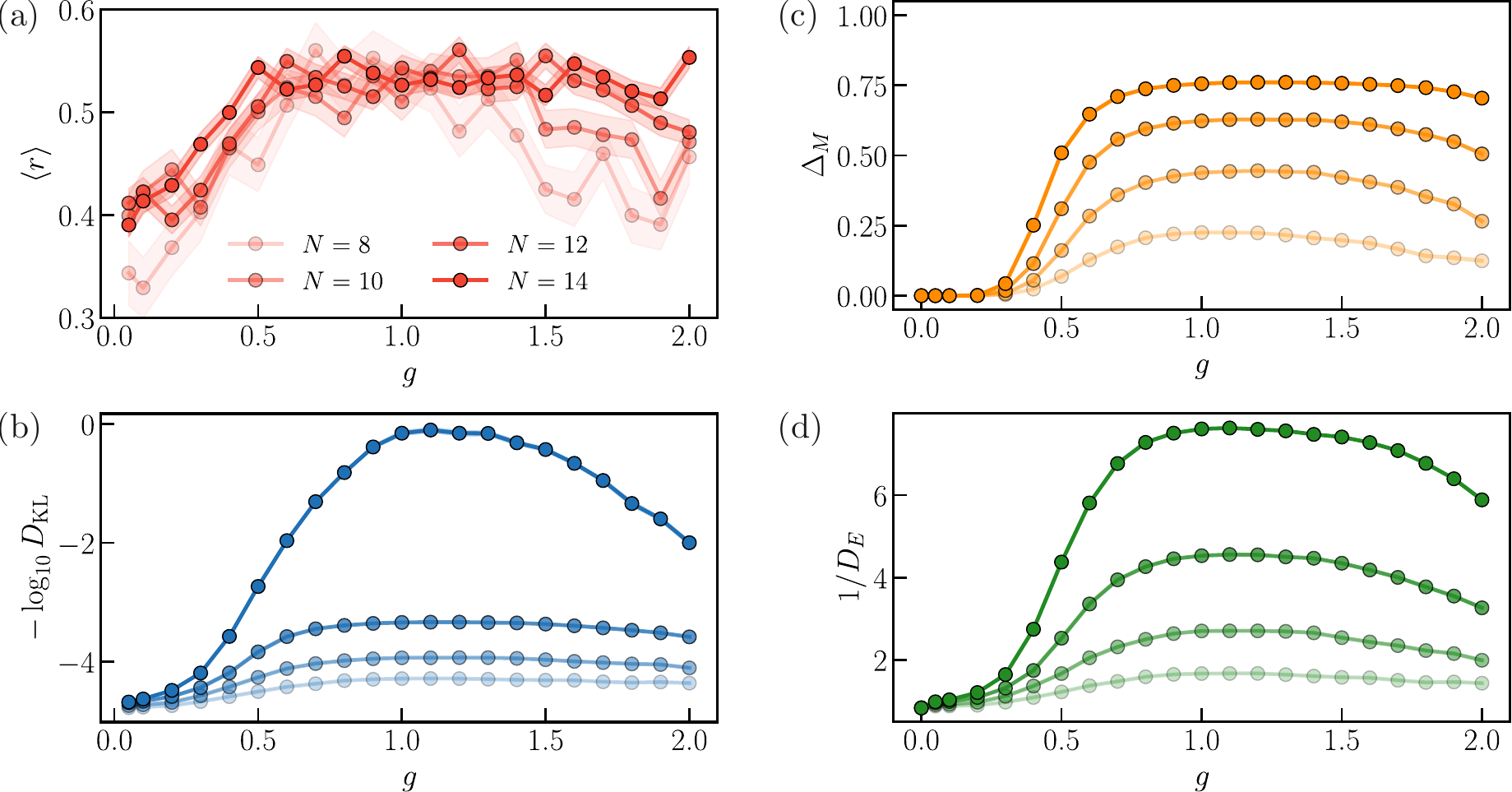}
    \caption{
    \edit{
    \textbf{Scaling of Ergodicity Metrics with System Size.}
    To complement the results in Fig.~\ref{fig:ergodicity maps}, we plot $h = 0.3$ cuts for each metric at system sizes $N = 8$, $10$, $12$, and $14$ for microcanonical ensembles centered at $E = 0$ with $100$, $200$, $400$, and $600$ eigenstates respectively.
    In each case, we also plot the standard error of each metric as a ribbon, computed in the same way as in Fig.~\ref{fig:snr cuts}.
    (a) As expected, $\langle r \rangle$ is dominated by fluctuations at each system size, whereas the other metrics (b-d) show barely any fluctuations at all.
    We attribute the large jump in the behavior of $- \log_{10} D_{\rm KL}$ from $N = 12$ to $14$ to the alignment of the system's entanglement distribution with that of the Bianchi-Dona distribution \cite{Rodriguez-Nieva_2024}.
    }
    }
    \label{fig:metric cut scaling}
\end{figure*}

\edit{
Given the large fluctuations in the level spacing ratio in Figs.~\ref{fig:ergodicity maps} and \ref{fig:snr cuts} when computed for 600 eigenstates, one might wonder whether computing $\langle r \rangle$ over the entire energy spectrum would improve its SNR and also allow it to isolate the maximally ergodic region found by the other metrics.
In this appendix, we investigate this possibility, with the results shown in Fig.~\ref{fig:full level spacing}.
First, Fig.~\ref{fig:full level spacing}(a) shows the same map as in Fig.~\ref{fig:ergodicity maps}(a) with the level spacing ration averaged over the entire spectrum rather than a microcanonical ensemble.
While the fluctuations have been substantially reduced, no structure emerges in the map within the dark red ergodic regime.
}

\edit{
To get a better understanding of whether this value of $\langle r \rangle$ can quantify ergodicity continuously, we also plot cuts of the data in panel (b) for various system sizes along the $h = 0.3$ line.
We also include the standard error of the mean for these points, calculated in the same way as the values in Fig.~\ref{fig:snr cuts}(a).
Despite the reduced fluctuations from a larger ensemble of states, there is no clear delineation between a true ``signal'' in the SNR sense and fluctuations from eigenstate sampling in the ergodic regime $(g \gtrsim 0.5$).
While calculating the SNR would give a quantitative answer to whether any such signal exists, we no longer have a proxy for what the true signal is.
Indeed, our proxy for the true signal in Fig.~\ref{fig:snr cuts}(c) was the $N = 14$ data plotted in Fig.~\ref{fig:full level spacing}(b) due to its relatively suppressed fluctuations.
Because of this, we instead opt to zoom in on the data by plotting its absolute deviation from the value $\langle r \rangle_{\rm GOE} \approx 0.536$ predicted by the Gaussian Orthogonal Ensemble (GOE), the RMT ensemble that models the MFIM due to its time reversal symmetry \cite{Rodriguez-Nieva_2024,Atas_2013}.
We plot this value on a logarithmic scale in Fig.~\ref{fig:full level spacing}(c); however, on this scale it's clear that the point-to-point fluctuations in the metric are still dominated by the error bars represented by the red ribbon.
As such, our numerical evidence points to the level spacing ratio being far less sensitive to changes in ergodicity---even when averaged over the entire energy spectrum---due to its large fluctuations and low SNR.
}

\edit{
To also understand how the plots in Fig.~\ref{fig:snr cuts} behave as a function of system size, we plot $h=0.3$ cuts of these metrics in Fig.~\ref{fig:metric cut scaling} for $N = 8$, 10, 12, and 14.
As in the text, we choose a microcanonical ensemble of eigenstates centered at $E = 0$ with size 100, 200, 400, and 600 respectively.
Just as before, the level spacing suffers from large fluctuations that obscure any kind of signal one might find in the ergodic regime, while the other metrics vary smoothly and have much smaller fluctuations [each panel has ribbons denoting the standard error, but they are too small to see in panels (b-d)].
While the large jump in the behavior of $- \log_{10} D_{\rm KL}$ from $N = 12$ to 14 may seem odd, a similar jump was observed in Ref.~\cite{Rodriguez-Nieva_2024} when the numerical distribution of entanglement entropy finally overlapped with the RMT Bianchi-Dona prediction (see e.g. Fig.~6 in \cite{Rodriguez-Nieva_2024}).
Finally, it is worth noting that despite their high SNR, our metrics do show much less variation at $N=14$ than $-\log_{10} D_{\rm KL}$ does.
While this is not an issue at the current system sizes, since we can safely interpret small changes in our metrics as a signal due to the much smaller noise, it is unclear whether this will impact the metrics' sensitivity as $N$ further increases and experimental imperfections are introduced.
We leave these questions to future work.
}

\section{Error Bounds from Shot Noise}
\label{app:noise error bounds}

We now analyze how the additive error $\epsilon$ on any individual matrix element of $M$ propagates to an error on $D_E$.
We will consider the case of an eigenstate of an ergodic Hamiltonian with no other conserved quantities for large $N$ such that $\Delta_M$ is close to 1.
Because this implies that the estimated Hamiltonian is robust to errors, we will project out this component of our operator space and only consider the effect of shot noise on the spectrum of variances near 1.
Modifying our previous notation to avoid confusion regarding $\epsilon$, we write $M = M_0 + \mathcal{E}$, where $M_0$ is the eigenstate's covariance matrix, $M$ is the measured covariance matrix, and $\mathcal{E}$ encapsulates the error from shot noise.
We first note that 
\begin{align}
    ||\mathcal{E}||_2 = \sqrt{\sum_{ij}\mathcal{E}_{ij}^2} \leq \sqrt{\sum_{ij} \epsilon^2} = N_L \epsilon
\end{align}
due to the additive error from shadow tomography.
We then project out the Hamiltonian from our matrices using $P \equiv I - |H)(H|$:
\begin{align}
    P M P = P M_0 P + P \mathcal{E} P ,
\end{align}
where $||P \mathcal{E} P||_2 \leq ||\mathcal{E}||_2 \leq N_L \epsilon$.
Now, subtract the identity matrix from both sides and apply the triangle inequality:
\begin{align}
    ||P M P - I||_2 \leq ||P M_0 P - I||_2 + ||P \mathcal{E} P||_2 .
\end{align}
Note that $||P M_0 P - I||_2 = \sqrt{\sum_{a \neq 0} (\sigma_a^2 - 1)^2}$, where $\sigma_a^2$ are the eigenvalues of $M_0$, and is therefore equal to $\sqrt{N_L-1} D_E$. 
Similarly, $||P M P - I||_2$ well approximates $ \sqrt{N_L-1} D_E$ for the matrix $M$, with the error coming from the fact that the estimated Hamiltonian is not exactly $H$ and therefore is not completely removed from the sum.
We will therefore refer to $||P M P - I||_2$ as $\sqrt{N_L-1} D_E'$ to avoid confusion.
Given this notation, our inequality becomes
\begin{align}
    D_E' - D_E \leq \frac{N_L}{\sqrt{N_L-1}} \epsilon .
\end{align}
The triangle inequality also implies that $D_E - D_E'$ is upper bounded in the same way, giving
\begin{align}
    |D_E' - D_E| \leq \frac{N_L}{\sqrt{N_L-1}} \epsilon,
\end{align}
suggesting that a target additive error $\epsilon_D$ in $D_E$ should require an additive error $\epsilon$ in the matrix elements of $M$ that scales like $O(1/\sqrt{N})$.

\end{document}